\documentclass[prb,amssymb,twocolumn,showpacs,supersriptaddress,floatfix]{revtex4-1}
\usepackage{amssymb}
\usepackage{amsmath}
\usepackage{graphicx}
\usepackage{bm}
\usepackage{caption}
\usepackage{subcaption}
\usepackage{bm}
\usepackage{natbib,hyperref}
\usepackage{hyperref}
\usepackage{color}
\hypersetup{colorlinks=true, 
    linkcolor=red,          
    citecolor=magenta,        
    filecolor=gree,      
    urlcolor=blue           
}

\graphicspath{{images/}}

\begin{document}


\title{Driving controlled entanglement in coupled flux qubits}

\author{Ana Laura Gramajo, Daniel Dom\'{\i}nguez and Mar\'{\i}a Jos\'{e} S\'{a}nchez}

\affiliation{Centro At{\'{o}}mico Bariloche and Instituto Balseiro, 8400 San Carlos de Bariloche, Argentina.\\
}




\date{\today}%
\begin{abstract}
We study the manipulation of quantum entanglement by periodic external fields. As an entanglement measure we compute numerically the concurrence of two flux qubits coupled  inductively and/or capacitively, both driven by a dc+ac magnetic flux.
Also we find an analytical  lower bound for the concurrence, where the dominant terms correspond to   the concurrence in the Floquet states.
 We show that it is possible to create or destroy entanglement in a controlled way by
 tuning the system at or near multiphoton resonances. We find that when the 
 driving term of the Hamiltonian does not commute with the qubit-qubit interaction term, the control of the entanglement induced by the  driving field is more robust in parameter space. This implies that capacitively coupled two flux qubits are  more convenient for controlling entanglement through ac driving fluxes.
\end{abstract}

\maketitle

\section{\label{sec:I}Introduction}
Entanglement is a central resource for  quantum  protocols that exploits  the  non local correlations absent 
in its  classical analogues. 

Among the rich variety of qubits that  have been  explored as candidates for  quantum
computation, solid state  superconducting devices based on Josephson junctions circuits are promising
due to  their  microfabrication techniques and downscalability. 
\cite{orlando,friedman,vanderwal,martinis}

To implement a quantum algorithm, one must be able to entangle  qubits
by means of  an  interaction term  in the Hamiltonian
describing a two qubit system. 
For  superconducting
flux qubits \cite{orlando,oliver}, the natural interaction is between
the magnetic fluxes, providing a   coupling through their mutual
inductance.\cite{orlando,oliver, majer} On the other hand, for phase or charge qubits the dominant coupling is essentially capacitive. \cite{strauch}

Although coupled superconducting  qubits have been experimentally realized \cite{izmalkov,grajcar,majer}, the generation and control of  entanglement can be  quite complicated and demanding, even requiring  sequences of single- and two-qubit operations. 
 Along this line, pulse sequences 
have been implemented  for several superconducting
qubits with fixed interaction energies \cite{yamamoto, strauch}. However,
entangling operations can be much more efficient if the
interaction can be varied and, ideally, turned off during
parts of the manipulation. Some tunable  coupling schemes have been proposed, \cite{mooij, nori1, wendin,plourde} but  these approaches failed in
controlling  the coupling entirely.

Alternatively,  engineering  selection rules of transitions among different energy levels is a possible strategy for coupling and decoupling superconducting qubits. \cite{harrabi}
Following this idea, a method to create artificial selection rules- by suppressing and/or exciting specific transitions- has been developed for a pair of superconducting flux  qubits \cite{mooij}, simultaneously driven by a single resonant frequency with different amplitudes and phases.\cite{degroot}

Our main goal  is to   study the possibility of manipulating  entanglement between two flux qubits, by  external  driving fields of variable amplitude and fixed frequency non resonant with any specific transition.
The sensitivity of the energy levels of a flux qubit driven by an external (dc+ac) magnetic field, has been extensively studied in recent years.\cite{shevchenko,berns,rudner,izmalkov2,ferron2010} The experimental implementation  of  Landau-Zener-St\"uckelberg interferometry  \cite{oliver,olivervalenzuela} has become a tool to analyze quantum coherence under strong driving  and to access the multilevel structure of flux qubits, which exhibits several avoided crossings as a function of the magnetic flux.\cite{ferron2010}
In most  of these cases the  Floquet formalism has been employed to solve the system dynamics in terms of quasienergies and Floquet states. \cite{shirley,ferronprl,ferron2016}

In Ref.\onlinecite{sauer} it has been recently shown how to generate entangled Floquet states in weakly  interacting two level systems by controlling the amplitude  of the external driving field.
However Floquet states are not accessible experimentally. Here we  extend the study for arbitrary coupling beyond the weak interaction case,  and  for general pure states that can be  prepared as initial states,
like  the ground state of the two qubit system or states of the computational basis.
In particular we analyze how the strength and the kind of coupling between the two qubits affect the dynamics and the entanglement, considering different  static couplings for a given  microwave driving field configuration. 
 
 The system of work consists in two coupled flux qubits  driven by (the same) microwave field. As usual, each qubit is represented by a two level system \cite{sang,griffoni,nori2} and we focus here on the case of negligible dissipation and/or interaction with the environment.
  The two-qubits can be coupled  longitudinally or transversely. This corresponds respectively to the inductive (i.e.  by the mutual inductance) and capacitive (i.e. mutual capacitance) coupling  \cite{wendin}. As a measure 
of entanglement  we choose the \textit{concurrence} \cite{wootters} which increases monotonically from $0$ for non-entangled states to $1$, for maximally entangled states. 

Experimentally there are some  evidences of entanglement measurements in solid state physics, although the field is still immature. An advance in the ability to quantify the entanglement
between qubits was reported in Ref. \onlinecite{dicarlo}, with the implementation of a joint readout of 
superconducting qubits using a microwave cavity as a single
measurement channel that gives access to qubit correlations.
Using state tomography, in Ref. \onlinecite{shulman202}  the full density matrix of a two qubit system has been measured and the concurrence and the fidelity of the generated state determined, providing an experimental proof  of entanglement.
Additionally  there are  other proposals based on the  measurement of the ground state population of two copies of a bipartite system \cite{romero}, that could give direct access to the  concurrence  for pure states.

The paper is organized as follows. In Sec.\ref{sec:II} we present a brief description of  the system model for  two coupled  superconducting flux qubits and we introduce the concurrence in terms of Floquet states and  quasienergies. Section \ref{sec:III} is devoted to present the numerical and analytical results for the  concurrence taking into account different types of coupling between the two qubits, longitudinal and  transverse  respectively. We analyze the dependence of the concurrence on the parameters of the microwave field  and the coupling strengths. 
An analytical expression for a lower bound of the  concurrence in terms of the Floquet states is presented.
Finally in sec.\ref{conc} we summarize our results together with the conclusions and perspectives. 

\section{\label{sec:II} Concurrence for the two coupled flux qubits model}

The dynamics of two coupled flux qubits can be described by the global Hamiltonian \cite{akisato,satanin} 
\begin{equation}
  \hat{H}_{0}=-\frac{1}{2}\sum_{i=1}^{2}\left(\epsilon_{i}\sigma_{z}^{(i)} + \Delta_{i}\sigma_{x}^{(i)}\right)+\hat{H}_{12},
  \label{eq:1}
\end{equation} where $\epsilon_{i}$ is the detuning energy (which is proportional to the difference between the magnetic flux through the qubit and half the quantum of flux),  $\Delta_{i}$ is the tunnel splitting energy and $\sigma^{(i)}_{z},\sigma^{(i)}_{x}$ the Pauli matrices , with $i=1,2$ the index of each qubit. $\hat{H}_{12}$ is the coupling Hamiltonian, which for the  longitudinal and transverse couplings can be written respectively as:
\begin{equation}
  \begin{aligned}
     \hat{H}^{z}_{12}&=-\frac{J^{z}}{2}\sigma_{z}^{(1)} \otimes \sigma_{z}^{(2)},\\
     \hat{H}^{c}_{12}&=-\frac{J^{c}}{2}\left( (1-p)\sigma_{x}^{(1)} \otimes \sigma_{x}^{(2)} +  p\sigma_{y}^{(1)} \otimes \sigma_{y}^{(2)} \right),
  \end{aligned}
  \label{eq:2}
\end{equation} with $J^{z}$ and $J^{c}$ the correspondent coupling constants. As we already mentioned, it is possible to associate the longitudinal coupling to an inductive coupling \cite{wendin,majer}, since $J^{z}=\pm MI_{1}I_{2}$ can be written in terms of the  mutual inductance $M$ and  the qubits currents  $I_{1,2}$. For $J^{z}<0$ ($J^{z}>0$) the coupling is  antiferromagnetic (ferromagnetic).  
For the transverse  coupling Hamiltonian,  $J^{c}$ is associated with the mutual capacitance between qubits. \cite{wendin,nori1}
 We here introduce  a mixing factor $p\in\mathbb{R}$ to take into account different proposals of transverse coupling \cite{wendin}.

To study the dynamics in the presence of driving fields,  we replace as usual $\epsilon_{i} \rightarrow \epsilon_{i}(t) = \epsilon_{i} + f(t)$ \cite{satanin,sauer,nori2,akisato} where $f(t)=A\cos(\omega t)$ is the microwave field (magnetic flux)  of amplitude $A$ and  frequency  $\omega$ applied to each qubit.  

The resulting Hamiltonian  is thus periodic in time, $\hat{H}(t)=\hat{H}(t+T)$ with period $T=2\pi/\omega$. According to the Floquet theorem \cite{shirley,sang,griffoni}, the solution of the Schr\"odinger equation   can be spanned in the Floquet basis $\{|u_{\alpha}(t)\rangle\}$ as $|\Psi (t)\rangle = \sum_{\alpha} a_{\alpha}(t_{0}) e^{-i\gamma_{\alpha}t/\hbar}|u_{\alpha} (t)\rangle$, with $\gamma_{\alpha}$ the quasienergies and $\alpha$ the index labeling the eigenstates of the time independent problem. For an initial condition $|\Psi(t_{0})\rangle$  at time $t_{0}$, we define the coefficients $a_{\alpha}(t_{0})=\langle u_{\alpha}(t_{0})|\Psi(t_{0})\rangle$.
 The time-evolution for a Floquet state is given by  $(H(t) -i\hbar \frac{\partial}{\partial t})|u_{\alpha} (t)\rangle =  \gamma_{\alpha}|u_{\alpha} (t)\rangle$, and the satisfy $|u_{\alpha} (t+T)\rangle=|u_{\alpha} (t)\rangle $. Therefore  after expanding the time periodic Floquet states in the Fourier basis, $|u_{\alpha} (t)\rangle=\sum_k e^{ik\omega t}|u_{\alpha} (k)\rangle $ the time-dependent problem is reduced  to a time-independent eigenvalue problem. For more details see Appendix \ref{ap:A}.


An entanglement measure quantifies the degree of quantum correlations present in a given quantum state.
In the case of  pure states $|\Psi(t)\rangle$ a useful quantity is  the concurrence  \cite{wootters}, 
\begin{equation}
  C(t,t_{0})=|\langle \Psi(t)|^{*}\sigma_{y}^{(1)}\otimes\sigma_{y}^{(2)}|\Psi(t)\rangle|,
\label{eq:3}
\end{equation} 
that  goes from $0$ for non-entangled states, to $1$ for maximally entangled states.  
Notice that  Eq.(\ref{eq:3}) depends implicitly  on the initial time $t_{0}$ through  $|\Psi(t)\rangle$.

Using the extended Floquet basis in Fourier space $\{|u_{\alpha}(k)\rangle\}$ with $k\in\mathbb{Z}$,
Eq.(\ref{eq:3})  can be written as 
\begin{equation}
  C(t,t_{0})= \lvert\sum_{\alpha\beta kk'qq'}\tilde{C}_{\alpha\beta}(k,k')f_{\alpha\beta}(q,q')e^{-i\varphi_{\alpha\beta}^{kk'qq'}(t,t_{0})}\arrowvert,
  \label{eq:4}
\end{equation} where $\varphi_{\alpha\beta}^{kk'qq'}(t,t_{0})= (\gamma_{\beta}+\gamma_{\alpha} -(k'+k)\omega)t-(\gamma_{\beta}+\gamma_{\alpha} -(q'+q)\omega)t_{0}$, with $\gamma_{^{\alpha}_{\beta}}$ the quasienergies,  $\tilde{C}_{\alpha\beta}(k,k')= \langle u_{\alpha}(k)|^{*} \sigma_{y}^{(1)} \otimes \sigma_{y}^{(2)}  |u_{\beta} (k') \rangle$ and $f_{\alpha\beta}(q,q')= a_{\alpha}(q)a_{\beta}(q')$, with $a_{^{\alpha}_{\beta}}(^{q}_{q'})=\langle u_{^{\alpha}_{\beta}}(^{q}_{q'}) | \Psi(t_{0}) \rangle$
(see Appendix \ref{ap:A}).
 Under general conditions,  the initial time should be averaged out, thus
 we  compute the time-averaged of Eq.(\ref{eq:4}) over $t_{0}$, 
\begin{equation}
  \overline{C}(t)=\frac{1}{T}\int_{0}^{T}dt_{0}\,C(t,t_{0}),
\label{eq:5}
\end{equation}  
 which  still presents an oscillating behaviour with time $t$ (see Fig. \ref{fig:2}), typical for time dependent driven systems. However and  unlike the occupation probability, the concurrence is not a periodic function of the driving period, as can be easily checked from its definition, Eq. (\ref{eq:4}).
In order to eliminate the dependence on time $t$ we perform an additional time average, obtaining
\begin{equation}
  \overline{\overline{C}}=\lim_{ t^{\prime}\rightarrow \infty}\frac{1}{t^\prime}\int_{0}^{t^\prime}dt\,\overline{C}(t).
\label{eq:6}
\end{equation}

\section{\label{sec:III}Results}

To solve  the dynamics  for both types of  couplings (see Eq.(\ref{eq:2})), we use the Floquet formalism described in the previous section. 

After computing numerically the Floquet states and  quasienergies,  we calculate the concurrence using Eq.(\ref{eq:4}) and the  respective averages over $t_{0}$ and $t$, given
in Eqs. (\ref{eq:5}) and (\ref{eq:6}). 
In the Appendix \ref{ap:B} we derive an  analytical expression $C_{I}$, which is  a lower bound of the time-averaged concurrence $\overline{\overline{C}}$, whose interpretation in terms of Floquet states  will be discussed below.

Along this work, we fix  $\epsilon_{i}=\epsilon_{0}$, $\forall i=1,2$, and  choose without loss
of generality, $\Delta_{1}/\omega=0.1$ and $\Delta_{2}/\omega=0.15$. We take $\hbar=1$ and 
energy scales are normalized by $\omega$. 

\subsection{\label{subsec:IIIA}Longitudinal coupling}

We start  with the results for the inductive coupling Hamiltonian in the presence of driving,
\begin{equation}
  \hat{H}_{I}(t)=-\frac{1}{2}\sum_{i=1}^{2}\left(\epsilon_{i}(t)\sigma_{z}^{(i)} + \Delta_{i}\sigma_{x}^{(i)}\right) - \frac{J^{z}}{2}\sigma_{z}^{(1)}\otimes\sigma_{z}^{(2)}.
  \label{eq:7}
\end{equation}


\begin{figure}[!htb]
\begin{subfigure}[ht!]{0.5\textwidth} 
\includegraphics[scale = 0.25]{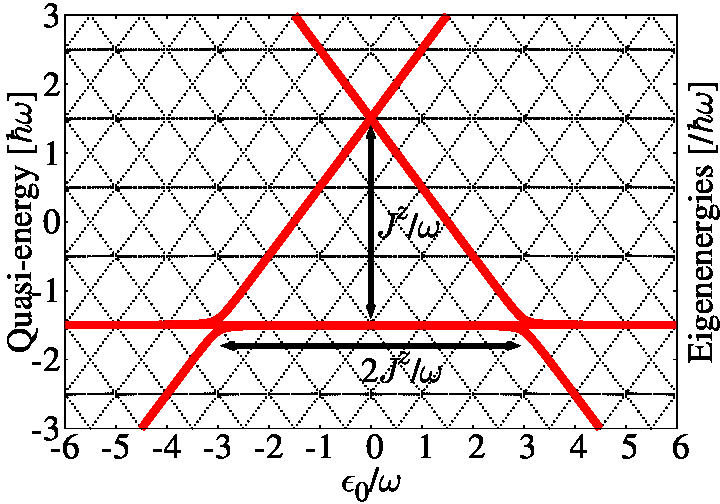}
	    \caption {$J^{z}/\omega=-3$}
	    \label{fig:1a}
        \end{subfigure}
        \
        \begin{subfigure}[ht!]{0.5\textwidth} 
	    \includegraphics[scale = 0.25]{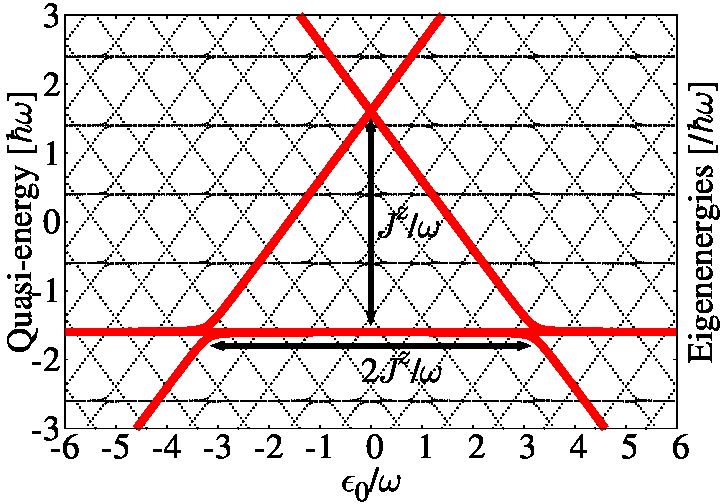}
	    \caption {$J^{z}/\omega=-3.2$}
	    \label{fig:1b}
        \end{subfigure}
  \caption{Eigenenergies  in the absence of driving,  $A/\omega=0$ (red lines) and quasienergies for $A/\omega=3.8$ (black lines) as a function of $\epsilon_{0}/\omega$, for the coupling strengths $J^{z}/\omega = -3$ (a) and $J^{z}/\omega = -3.2$ (b) (see Hamiltonian Eq.(\ref{eq:7})). Parameters are  $\Delta_{1}/\omega=0.1$ and $\Delta_{2}/\omega=0.15$.}
\label{fig:1}
\end{figure}

Fig.\ref{fig:1} shows as a function of the  detuning $\epsilon_{0}/\omega$ and  for the longitudinal coupling strengths $J^{z}/\omega=-3$ and $J^{z}/\omega = -3.2$, the eigenenergies $E_{i}$, $i=0,...,3$ for the time indepent Hamiltonian (red lines), and in black the quasienergies for the driven Hamiltonian Eq.(\ref{eq:7}) for  $A/\omega=3.8$. We work with the (antiferromagnetic) coupling $J^{z}<0$, which reduces the energy of states $|01\rangle,|10\rangle$ while it increases the energy of  states $|00\rangle,|11\rangle$, with $\mathcal{E}=\{|00\rangle,|01\rangle,|10\rangle,|11\rangle\}$  the computational basis in the product space of the two qubits. 

The quasienergies display avoided crossings (quasidegeneracies), where the Floquet states are strongly mixed.  These quasidegeneracies will  play a central role in the structure of the concurrence as we will
discuss in the following.
In the limit of $\Delta_{i}/\omega\rightarrow 0$, the quasidegeneracies transform in exact crossings of quasienergies,  giving the resonance condition $\gamma_{\alpha}-\gamma_{\beta}= n \omega$, $n\in\mathbb{Z}$. \cite{sang}

For the present  case of longitudinal coupling the quasienergies of the two qubit system computed analytically  
for $\Delta_{i}/\omega\rightarrow 0$ using the  Van Vleck  nearly degenerate
perturbation theory \cite{sang} are $\gamma_{\alpha}\sim  \pm \epsilon_{0} + J^{z}/2 + m \omega$ and the (quasi) degenerate pair
$ -J^{z}/2 + m \omega$
 with $m\in\mathbb{Z}$. As, for $\Delta_{i}/\omega\rightarrow 0$,
the driving $\epsilon(t)$ and the static coupling Hamiltonian are both  proportional to  ${\sigma_{z}}$, 
 the location of the quasi crossings in the spectrum of quasienergies are replicas (in $\pm m \omega$)   of the quasi crossings of  the  static spectrum (see Fig.\ref{fig:1}).
The resonance conditions are thus satisfied respectively for $2 \epsilon_{0}\sim n \omega$ and $\epsilon_{0}\pm J^{z} \sim n \omega$,  and correspond to  multiphoton processes where the population probability 
is modulated by the zeros of the Bessel functions of order $n$, $J_{n}(A/\omega)=0$.\cite{satanin,sang} 
Notice that while the first condition gives half integer and integer values of $\epsilon_{0}/\omega$, 
 the second one depends on $J^{z}/\omega$.
 For integer values of $J^{z}/\omega$ the quasidegeneracies are located at integer values of $\epsilon_{0}/\omega$, as can be clearly seen from  Fig.\ref{fig:1a}.
However, for arbitrary $J^{z}/\omega$,   quasi degeneracies also appear for  values of $\epsilon_{0}/\omega$ which are neither integer nor half integers (see  Fig.\ref{fig:1b}).
\begin{figure}[!htb]
\includegraphics[scale = 0.35]{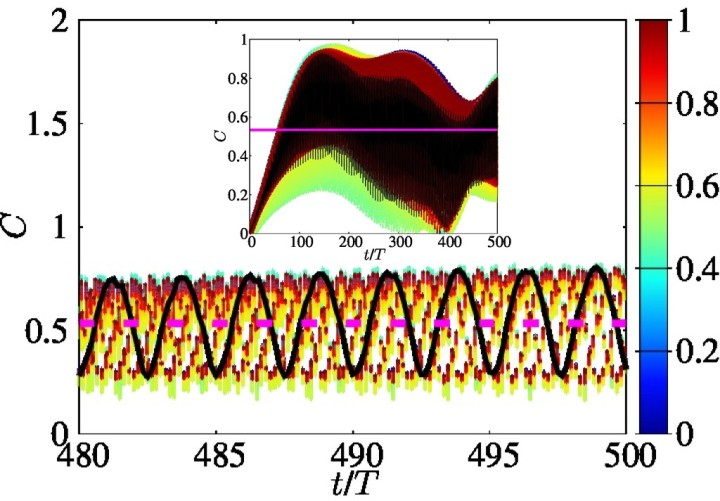}
\caption{Plots of the concurrence ${C(t, t_{0})}$ for $\epsilon_{0}/\omega = 4$  and $A/\omega=3.8$,  as a function of the normalized time $t/T$ for $300$ different initial times $t_{0}/T$, represented on the  vertical colour bar. The initial state is  $|\Psi(t_{0})\rangle$. In  black line  $\overline{C(t)}$ is plotted and in magenta,  its average $\overline{\overline{C}}$. The inset show the plot in the range $t/T\in[0,500]$, while the main figure shows a detail in an interval $t/T\in[480,500]$. Results correspond to inductive coupling with $J^{z}/\omega = -3$, see Eq.(\ref{eq:7}). The other qubits parameters are the same  than in Fig.\ref{fig:1}.}
\label{fig:2}
\end{figure}

Fig.\ref{fig:2} displays the concurrence $C(t,t_{0})$ as a function of the normalized time $t/T$ calculated for the initial condition $|\Psi(t_{0})\rangle= e^{-i E_{0} t_{0}} |E_{0}\rangle$, which was chosen as the ground state of the time independent Hamiltonian  (with eigenvalue $E_{0}$) for  detuning energy $\epsilon_{0}/\omega=4$ and static coupling strength $J^{z}/\omega=-3$ (see Fig.\ref{fig:1}).
The driving amplitude chosen is $A/\omega=3.8$. $C(t,t_{0})$ was computed for around $300$ different initial times with $t_{0}/T\in[0,1]$. Notice that due to the different  initial times, that induce a different initial phase in the microwave field, the  curves are shifted.
After averaging over $t_{0}$,  $\overline{C}(t)$ (black line),
turns out to be a smoother function whose time average  $\overline{\overline{C}}$, results independent on time as expected. In the following we will focus  on $\overline{\overline{C}}$,
 which  contains the time independent contributions to entanglement dominant for long times.

\begin{figure}[!htb]
          
        \begin{subfigure}[ht!]{0.5\textwidth} 
	    
	    \includegraphics[scale = 0.25]{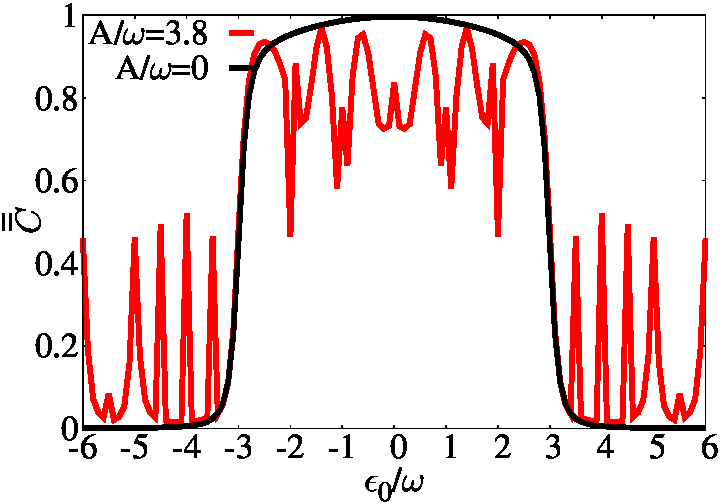}
	    \caption {}
	    \label{fig:3a}
        \end{subfigure}
        \
        \begin{subfigure}[ht!]{0.5\textwidth} 
	    
	    \includegraphics[scale = 0.25]{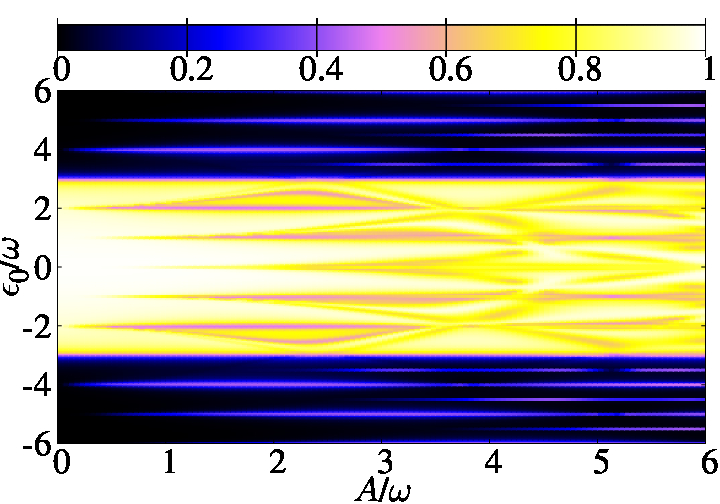}
	    \caption {}
	    \label{fig:3b}
        \end{subfigure}
 \caption {(a) Plots of $\overline{\overline{C}}$ versus $\epsilon_{0}/\omega$ for  $A/\omega=0$ (black line) and $A/\omega=3.8$ (red line). (b) Colour map of  $\overline{\overline{C}}$ versus $\epsilon_{0}/\omega$ and $A/\omega$. In both plots  the initial condition corresponds
 to the ground state  for the correspondent $\epsilon_{0}/\omega$. $J^{z}/\omega = -3$ is the coupling  strength (see Eq.(\ref{eq:7})) and other qubits parameters are the same as in Fig. \ref{fig:2}.}
\label{fig:3}
\end{figure}

In Fig.\ref{fig:3a} we show  $\overline{\overline{C}}$ as a function of $\epsilon_{0}/\omega$. 
 The initial state is chosen as the ground state for the correspondent  $\epsilon_{0}/\omega$, and we keep the same values than in the Fig.\ref{fig:2} for the  other parameters. In the absence of driving, $A/\omega=0$ (black line), the ground state is entangled for  detuning energies satisfying $|\epsilon_{0}/\omega|\lesssim |J^{z}/\omega|=3$, where the concurrence takes values  close to $1$. In particular, for $\epsilon_{0}=0$, the ground state is the Bell's state $(|01\rangle + |10\rangle)/\sqrt{2}$, which is known to be a maximally  entangled state.\citep{wootters}
On the other hand, for values of detuning energy  $|\epsilon_{0}/\omega|> |J^{z}/\omega|=3$ the ground state is almost disentangled, i.e. for large values of $\epsilon_{0}$ the ground state is asymptotically a separable state of the computational basis, corresponding to $|00\rangle$ for $\epsilon_ {0}\gg 0$ and $|11\rangle$ for $\epsilon_ {0}\ll 0$, see Fig.\ref{fig:1}. 

When the driving is turn on ($A/\omega=3.8$), a pattern of resonances is clearly visible in $\overline{\overline{C}}$,  where entanglement is either  created or destroyed.
For $|\epsilon_{0}/\omega|> |J^{z}/\omega|=3$, where the initial condition corresponds to a separable state, we see that it is possible to generate  entanglement in a controlled way around a given resonance. Otherwise entanglement is reduced.
 
The  positions of the resonances in $\overline{\overline{C}}$ are determined 
from the already mentioned conditions: $2 \epsilon_{0} /\omega\sim  n$ and $\epsilon_{0} /\omega + J^{z}/\omega \sim  n$, with  $n\in\mathbb{Z}$.  Therefore is around a (quasi) degeneracy  where the Floquet states  can be strongly mixed given rise to significant deviations
in the  behaviour of the concurrence compared to the undriven case.

Notice that in Fig.\ref{fig:3a}, the resonances are at integer and half integer values of $\epsilon_{0}/\omega$, since   $J^{z}/\omega=-3$.

In  Fig.\ref{fig:3b}, we plot   $\overline{\overline{C}}$ as a function of  $\epsilon_{0}/\omega$ and $A/\omega$, for $J^{z}/\omega=-3$. For a particular  multiphoton resonance,
the concurrence is modulated by the driving amplitude, where full (or partial) recovery of the initial entanglement is possible. 

\begin{figure*}[!htb]
          
        \begin{subfigure}[ht!]{0.4\textwidth} 
	    \includegraphics[scale = 0.25]{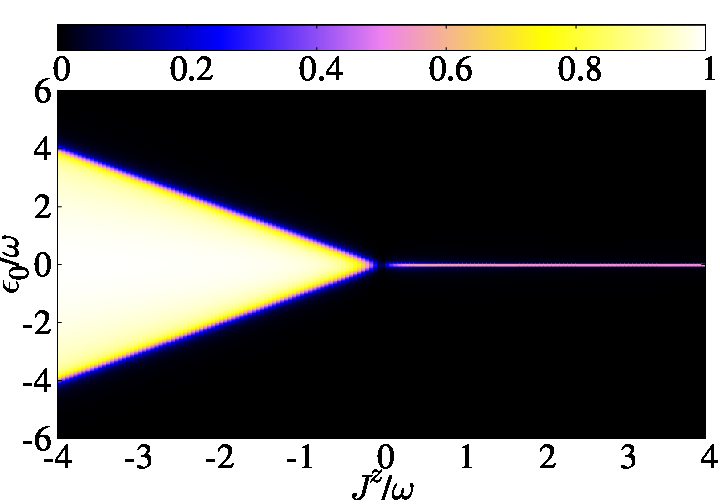}
	    \caption {$A/\omega=0$}
	    \label{fig:4a}
        \end{subfigure}
        \begin{subfigure}[ht!]{0.4\textwidth} 
	    
	    \includegraphics[scale = 0.25]{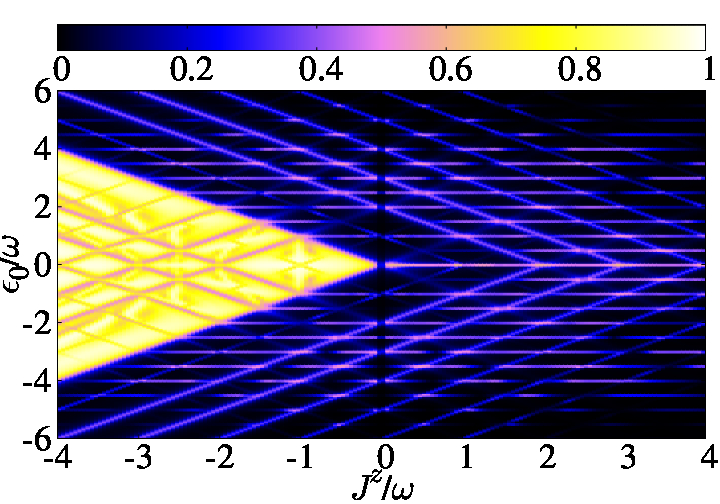}
	    \caption {$A/\omega=3.8$}
	    \label{fig:4b}
        \end{subfigure}
        \
        \begin{subfigure}[ht!]{0.4\textwidth} 
	    
	    \includegraphics[scale = 0.25]{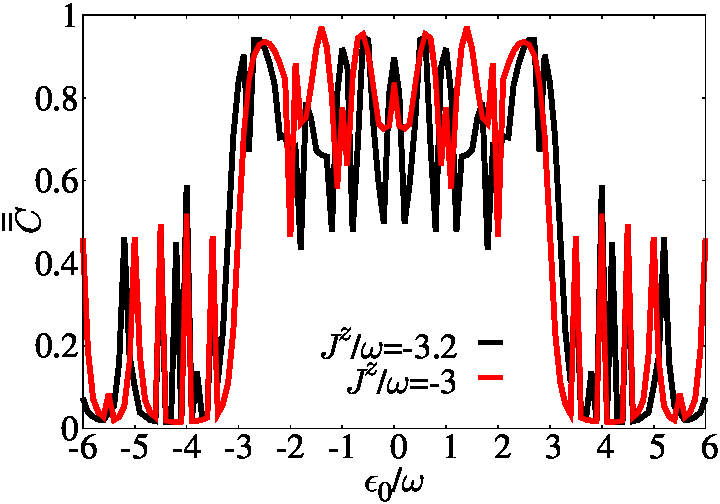}
	    \caption {$J^{z}/\omega <0$}
	    \label{fig:4c}
        \end{subfigure}
        \begin{subfigure}[ht!]{0.4\textwidth} 
	    
	    \includegraphics[scale = 0.25]{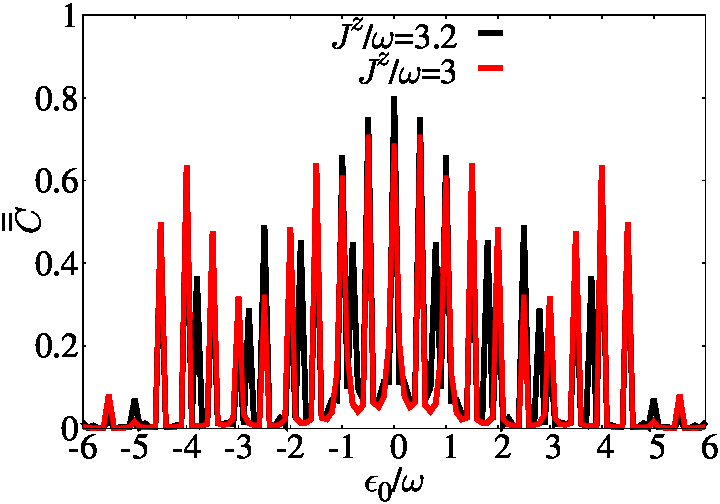}
	    \caption {$J^{z}/\omega>0$}
	    \label{fig:4d}
        \end{subfigure}
\caption {Colour map of  $\overline{\overline{C}}$ versus $J^{z}/\omega$ and $\epsilon_{0}/\omega$ for $A/\omega=0$ (a) and $A/\omega=3.8$ (b) respectively. 
 $\overline{\overline{C}}$ versus $\epsilon_{0}/\omega$ for $A/\omega=3.8$, for antiferromagnetic (c)  and ferromagnetic coupling strengths (d). See text for details.}
\label{fig:4}
\end{figure*}

So far, we  have studied the entanglement for a fixed value of the coupling strength. 
However in  practical implementations with flux qubits the intensity of the inductive coupling can be controlled, and it is interesting to analyse whether a different static coupling would induce qualitative changes in the above description, taking into account that the spectrum of quasienergies is sensitive to this change (see Figs.\ref{fig:1}).
In Fig.\ref{fig:4} we plot a map of  $\overline{\overline{C}}$ versus $J^{z}/\omega$ and $\epsilon_{0}/\omega$ for $A/\omega=0$ and 3.8, respectively, taking as the initial state  the ground state for the correspondent $\epsilon_{0}/\omega$ and $J^{z}/\omega$.
In the absence of  the microwave field (see Fig.\ref{fig:4a})  two
well separated behaviours are observed, corresponding to positive and negative values of $J^{z}$ respectively. For $J^{z}<0$  (antiferromagnetic coupling) the ground state is  entangled for $|\epsilon_{0}|<|J^{z}|$ as we already described, given rise to the triangular shaped  region in $\overline{\overline{C}}$. On the other hand, for the ferromagnetic coupling $J^{z}>0$, which  increases the energy of states $|01\rangle,|10\rangle$ and  decreases the energy of states $|00\rangle,|11\rangle$,   the ground state is entangled only for values $\epsilon_{0}\sim 0$, 
being approximately the Bell's state $(|00\rangle+|11\rangle)/\sqrt{2}$.
When the microwave field is on, $\overline{\overline{C}}$ exhibits
the structure of resonances where the entanglement is created or destroyed in a well controlled way (see Fig.\ref{fig:4b}) with resonances   located at half integer or integer values of  $\epsilon_{0}/\omega$ and others  located at  positions determined by the values of  
$J^{z}/\omega$, as we already mentioned. For integer values of $J^{z}/\omega$ the resonances  are
at integers   $\epsilon_{0}/\omega$, while for arbitrary real values   of $J^{z}/\omega$ they are respectively shifted to  non integer values of $\epsilon_{0}/\omega$ (notice  the straight lines forming the $>$-shaped pattern).  
These features are clearly observed in Figs.\ref{fig:4c} and \ref{fig:4d}, where  $\overline{\overline{C}}$ versus $\epsilon_{0}/\omega$ is plotted  for $J^{z}/\omega\lessgtr 0$ and $A/\omega=3.8$. 
  
In Appendix \ref{ap:B} we derive an expression for a lower bound $C_{I}$ of the  averaged concurrence $\overline{\overline{C}}$, that reads:
\begin{equation}
\begin{aligned}
 C_{I} = |\sum_{\alpha}\overline{\tilde{C}_{\alpha\alpha}(t)}\sum_{q}|a_{\alpha}(q)|^{2}|< \overline{\overline{C}} ,
\end{aligned}
  \label{eq:8}
\end{equation} where $a_{\alpha}(q)=\langle u_{\alpha}(q) | \Psi(t_{0}) \rangle$ has been already defined  and $\overline{\tilde{C}_{\alpha\alpha}(t)}$ is the time average of a \textit{Floquet preconcurrence} $\tilde{C}_{\alpha\alpha}(t)\equiv\langle u_{\alpha}(t)|^{*} \sigma_{y}^{(1)} \otimes \sigma_{y}^{(2)} |u_{\alpha} (t) \rangle$. 
\begin{figure}[!htb]
          
           \begin{subfigure}[ht!]{0.5\textwidth} 
	    
	    \includegraphics[scale = 0.25]{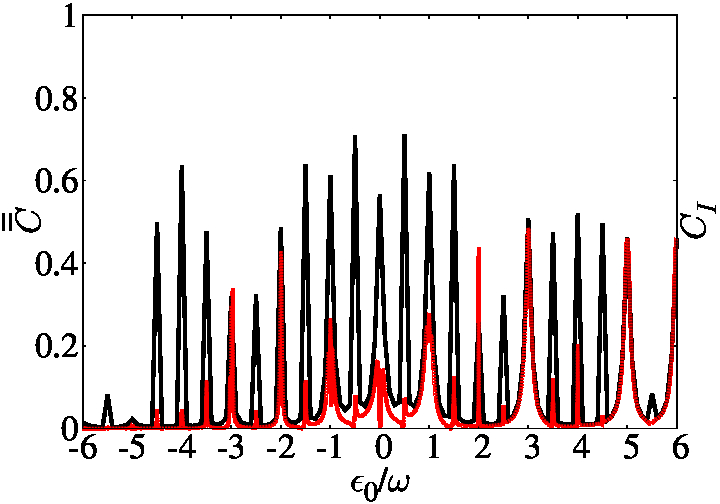}
	    \caption {}
	    \label{fig:5a}
        \end{subfigure}
        \begin{subfigure}[ht!]{0.5\textwidth} 
	    
	    \includegraphics[scale = 0.25]{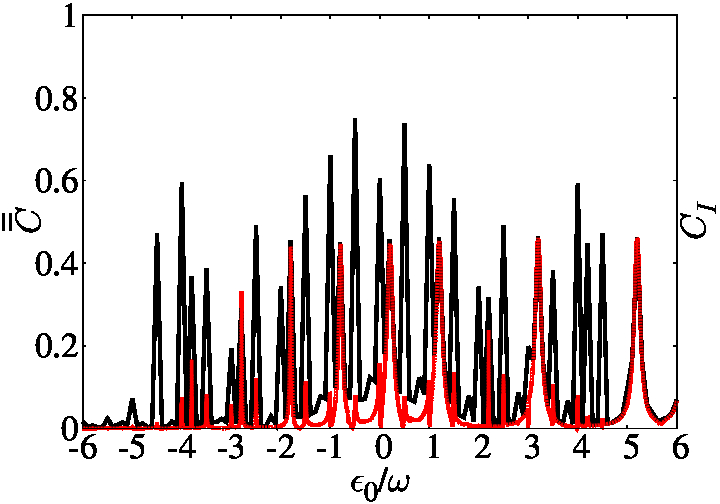}
	    \caption {}
	    \label{fig:5b}
        \end{subfigure}
       \caption{Plot of  $\overline{\overline{C}}$ (black line) and $C_{I}$ (red line) as a function of $\epsilon_{0}/\omega$. In both cases is  $A/\omega=3.8$ and the coupling strength $J^{z}/\omega = -3$ (a) and $J^{z}/\omega = -3.2$ (b).}
\label{fig:5}
\end{figure}
To obtain the above  expression for $C_{I}$ we have performed a rotating wave approximation disregarding fast oscillating terms in the concurrence, so we only consider the quasienergies that fulfilled the relation $\gamma_{\alpha} + \gamma_{\beta} = n\omega$, $n\in\mathbb{Z}$ (see Appendix \ref{ap:B}). 
In this way  $C_{I}$  is mainly governed by the \textit{Floquet preconcurrences}, each one weighted by the projection of the Floquet states on the initial condition. It should be noted that this expression is useful  when the initial state is  non entangled, since $C_{I}$ determines the minimal creation of entanglement. 
 
In  Fig.\ref{fig:5} we plot $\overline{\overline{C}}$ (black line)  and $C_{I}$  computed from Eq.(\ref{eq:8}) (red line), both as a function of $\epsilon_{0}/\omega$. We choose  the initial state $|00\rangle$, 
corresponding to a separable state (of the computational basis), and work with the  driven amplitude $A/\omega=3.8$. In Fig.\ref{fig:5a}  the coupling strength is  $J^{z}/\omega=-3$. 
 The position of the resonances at integer values of  $\epsilon_{0}/ \omega$ are well captured by the lower bound $C_{I}$, and  the agreement with $\overline{\overline{C}}$ is quite good.
 Notice however that for half integer values of  $\epsilon_{0}/ \omega$,  $\overline{\overline{C}}$ also exhibits resonances  that are quite attenuated in $C_{I}$.
This behaviour can be  understood taking into account that the stationary phase  condition employed to compute  $C_{I}$ involves the sum  $\gamma_{\alpha} + \gamma_{\beta}$ of pairs of  quasienergies, which gives  either  $\epsilon_0$ or $J^{z}$. Therefore for integer values of $J^{z}/\omega$ and $\epsilon_{0}/ \omega$,  the resonance conditions and the stationary phase  condition are both  satisfied.
 On the other hand, for half integer values of  $\epsilon_{0}/ \omega$ the stationary phase approximation is not fulfilled, and the resonances displayed in $\overline{\overline{C}}$ are
  dimmed  in $C_{I}$.
 
In Fig.\ref{fig:5b} we show a similar plot for  $ J^{z}/\omega= -3.2$, where 
 some  resonances in  $\overline{\overline{C}}$ are shifted to non integer values of $\epsilon_{0}/ \omega$, as we already discussed.
However, due to the stationary phase approximation, $C_{I}$ exhibits well defined resonances only at integer values of $\epsilon_{0}/ \omega$. As a consequence,   both curves are  horizontally displaced 
 respect to each other.

Owing  to the contribution of additional terms, $\overline{\overline{C}}$ displays  a background structure  not observed in  $C_{I}$ (see Appendix \ref{ap:B}).

\subsection{\label{subsec:IIIB}Transverse coupling}

In this section we focus on the  capacitive coupling, see Eq.(\ref{eq:2}), for which the
driven  time dependent Hamiltonian reads
\begin{equation}
\begin{aligned}
  \hat{H}_{II}(t)&=-\frac{1}{2}\sum_{i=1}^{2}\left(\epsilon_{i}(t)\sigma_{z}^{(i)} + \Delta_{i}\sigma_{x}^{(i)}\right) \\
  &- \frac{J^{c}}{2}((1-p)\sigma_{x}^{(1)}\otimes\sigma_{x}^{(2)}+p\sigma_{y}^{(1)}\otimes\sigma_{y}^{(2)}),
  \label{eq:9}
\end{aligned}
\end{equation} where  $p\in\mathbb{R}$ is a real number, and the other quantities have been already defined. 

\begin{figure}[!htb]
        \begin{subfigure}[ht!]{0.5\textwidth} 
	    \includegraphics[scale = 0.25]{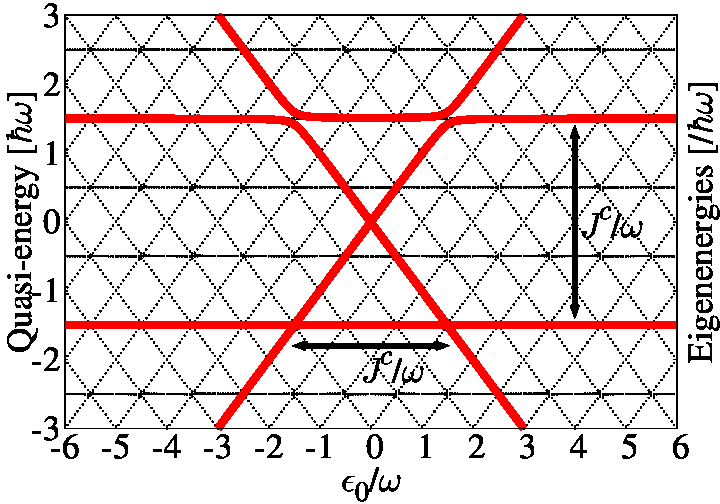}
	    \caption {$p=0.5$}
	    \label{fig:6a}
        \end{subfigure}
        \
        \begin{subfigure}[ht!]{0.5\textwidth} 
	    \includegraphics[scale = 0.25]{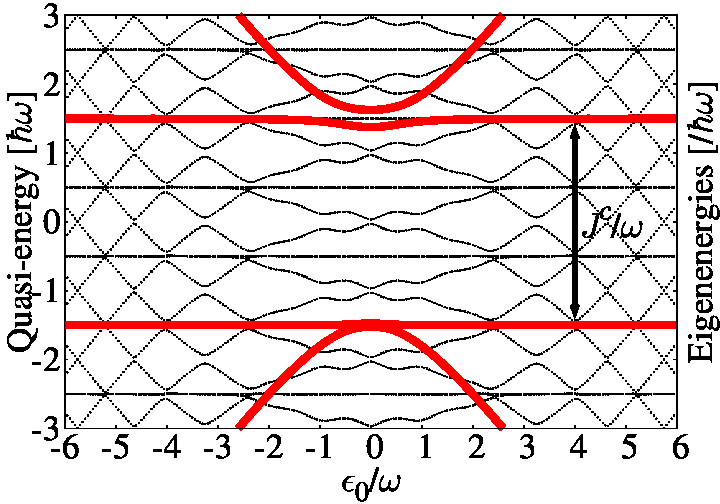}
	    \caption {$p=0$}
	    \label{fig:6b}
        \end{subfigure}
 \caption {Energy levels for $A/\omega=0$ (red lines) and quasienergies for $A/\omega=3.8$ (black lines) as a function of $\epsilon_{0}/\omega$ for $J^{c}/\omega = -3$, see Hamiltonian Eq.( \ref{eq:9}), with $p=0.5$ (a) and $p=0$ (b). 
 The  qubits parameters are the same as in previous figures.}
\label{fig:6}
\end{figure}

Figure \ref{fig:6} shows the eigenenergies  $E^{\prime}_{i}$, $i=0,...,3$, as a function of  $\epsilon_{0}/\omega$ for the static Hamiltonian   for   $p=0.5$ (Fig.\ref{fig:6a}) and $p=0$ (Fig. \ref{fig:6b}), in both cases for the coupling strength $J^{c}/\omega=-3$. 

 The coupling  Hamiltonian for $p=0.5$  
  represents the excitation exchange interaction  which  mixes the states $|01\rangle,|10\rangle$ breaking the former degeneracy present in the absence of coupling. 
For  $p=0$ the coupling Hamiltonian is $\hat{H}^{c}_{12}=J^{c}/2\sigma_{x}^{(1)}\otimes\sigma_{x}^{(2)}$, which includes the excitation exchange interaction and the magnetization-changing terms. \cite{sauer}
In this case  the coupling breaks the former degeneracy between $|01\rangle,|10\rangle$ but  also mixes the states $|00\rangle$ and $|11\rangle$, given rise to the spectrum  exhibited in Fig.\ref{fig:6b}.

Additionally we plot the quasienergies for the driven  Hamiltonian Eq.(\ref{eq:9})  in black lines, for the  amplitude $A/\omega=3.8$.
Notice that while  for $p=0.5$ the driving field 
along $\sigma_{z}$  and the coupling Hamiltonian commute, as it can be  checked from the properties of the Pauli matrices, for $p=0$ they do not.
As a consequence, the spectrum of quasienergies if Fig.\ref{fig:6b} has a non trivial structure. 

The quasienergies can be  obtained for arbitrary amplitudes
of the driving field for $p=0.5$  using perturbation theory for $\Delta_{i}/\omega\rightarrow 0$,  as we have done for the longitudinal coupling in Sec.\ref{subsec:IIIA}. For  the present case we get $\gamma_{\alpha} \sim \pm \epsilon_{0} + m \omega$ and $ \pm J^{c}/2+ m \omega$. Therefore the resonance conditions are satisfied for $\epsilon_{0}/\omega + J^{c}/2 \omega \sim n$, $2\epsilon_{0}/\omega\sim n$ and  $J^{c}/\omega \sim n$ (independent of the detuning), being quite similar to the inductive case.

However for the  $p=0$ case, the  additional limit $A/\omega\rightarrow 0$ has to be  taken in order to get   analytical expressions for the quasienergies $\gamma_{\alpha} \sim \pm \sqrt{\epsilon_{0}^{2} + (J_{c}/2)^{2}}+ m \omega $ and $\pm J^{c}/2 + m \omega$. Thus the resonance conditions are fulfilled, in this limit, for  $J^{c}/\omega\sim n$ (independent of the detuning), for values satisfying $(2 \epsilon_{0}/\omega)^{2} + (J^{c}/\omega)^{2} \sim n^{2}$ and  for $J^{c}/2\omega \pm \sqrt{ (\epsilon_{0}/\omega)^{2} + (J^{c}/2\omega)^{2}} \sim n$. The two latter relations  give rise to an intricate pattern of quasidegeneracies in  the spectrum of  Fig.\ref{fig:6b},  that will induce a  non trivial  behaviour in  the concurrence.
\begin{figure}[!htb]
 \includegraphics[scale = 0.25]{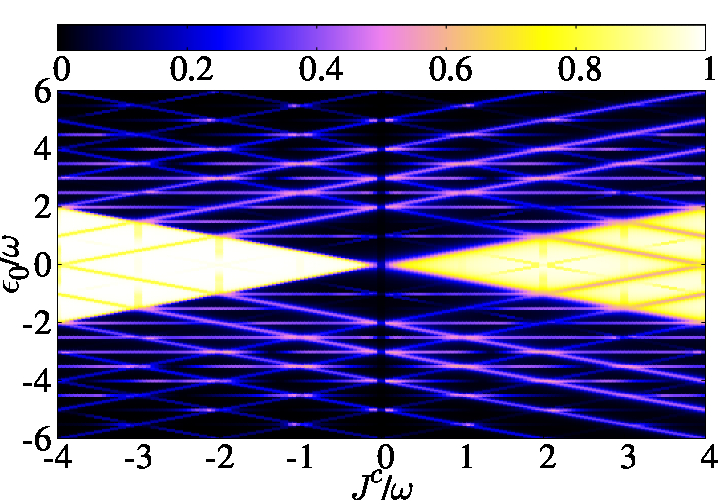}
\caption {Colour map of  $\overline{\overline{C}}$ versus $J^{c}/\omega$ and $\epsilon_{0}/\omega$ for  the capacitive coupling with $p=0.5$, see Hamiltonian Eq.(\ref{eq:9}).The driving amplitude is  $A/\omega=3.8$ (see text for more details).}
\label{fig:7}
\end{figure}

We start by analyzing the concurrence for  the symmetric case  $p=0.5$ where  the  response  is quite similar to the inductive coupling case previously studied.
In Fig. \ref{fig:7} we plot $\overline{\overline{C}}$ versus $J^{c}/\omega$ and $\epsilon_{0}/\omega$  when the driving amplitude is $A/\omega=3.8$ taking  as initial condition the ground state for each value of the detuning. It is evident that the  pattern of  resonances  where entanglement is created or destroyed resembles  the inductive case.

However, the \textit{symmetric transverse coupling} mixes the states $|01\rangle,|10\rangle$ independently of the sign of the coupling strength  $J^{c}$, given rise to similar ground states for both $J^{c}/\omega \lessgtr 0$, see Fig.\ref{fig:6a}. Consequently the concurrence exhibits a quite  symmetric  pattern of resonances  both in $J^{c}/\omega$ and $\epsilon_{0}/\omega$, unlike  the inductive studied in Sec. \ref{subsec:IIIA}.

For other values of the parameter $p\neq 0.5$,  we expect a non trivial behaviour of the concurrence due to the much richer  structure of the spectrum of quasienergies. 
\begin{figure*}[htb!]
        \begin{subfigure}[ht!]{0.4\textwidth} 
	    \includegraphics[scale = 0.25]{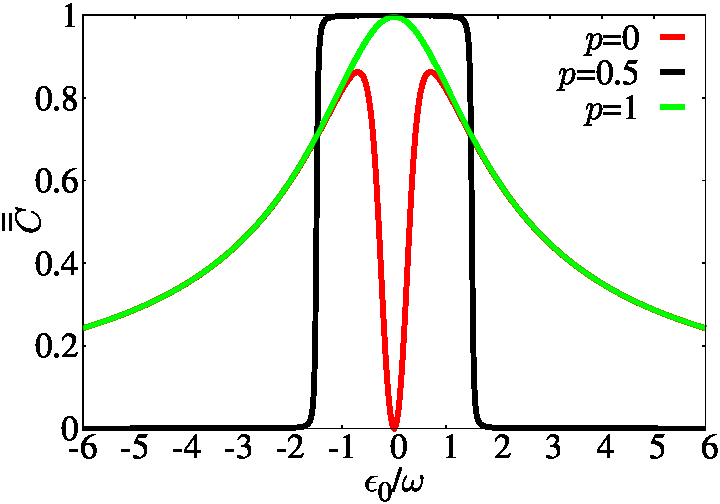}
	    \caption {$A/\omega=0$}
	    \label{fig:8a}
        \end{subfigure}
        \begin{subfigure}[ht!]{0.4\textwidth} 
	    
	    \includegraphics[scale = 0.25]{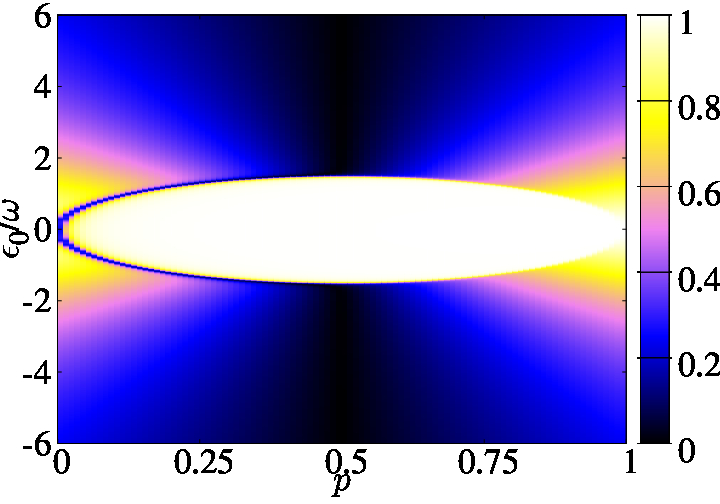}
	    \caption {$A/\omega=0$}
	    \label{fig:8b}
        \end{subfigure}
        \
        \begin{subfigure}[ht!]{0.4\textwidth}
	    
	    \includegraphics[scale = 0.25]{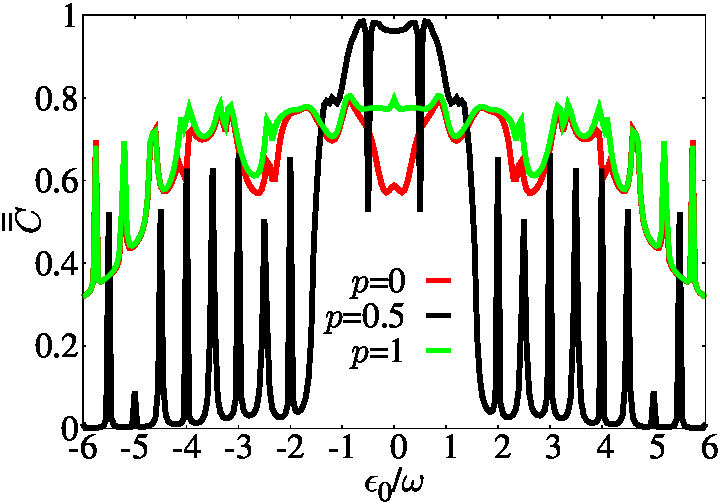}
	    \caption {$A/\omega=3.8$}
	    \label{fig:8c}
        \end{subfigure}%
        \begin{subfigure}[ht!]{0.4\textwidth}
	    
	    \includegraphics[scale = 0.25]{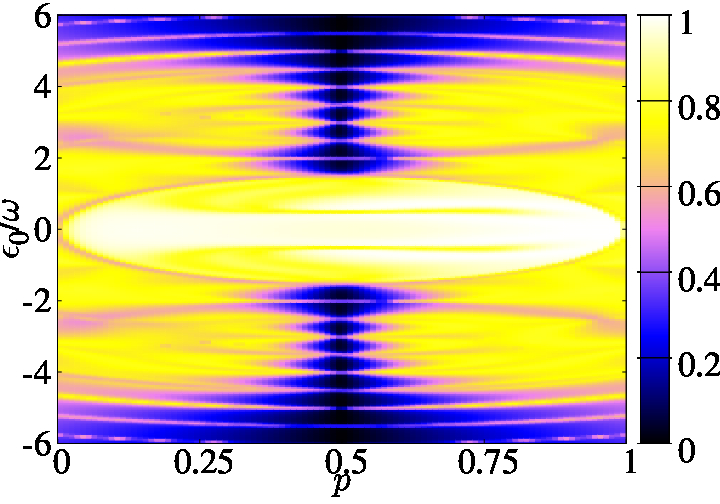}
	    \caption {$A/\omega=3.8$}
	    \label{fig:8d}
        \end{subfigure}%
        \caption{Plot of $\overline{\overline{C}}$  as a function of $\epsilon_{0}/\omega$  for $A/\omega=0$ (a) and $A/\omega=3.8$ (c).  $p=0$ (red line), $p=0.5$ (black line) and $p=1$ (green line).
        Colour map of $\overline{\overline{C}}$  as a function of $\epsilon_{0}/\omega$ and $p$ for $A/\omega=0$ (b) and $A/\omega=3.8$ (d) respectively. In all the cases the initial condition is the ground state $|E^{\prime}_{0}\rangle$. See text for more details.}
\label{fig:8}
\end{figure*}
\begin{figure*}[!htb]
  \begin{subfigure}[ht!]{0.4\textwidth} 
	   	    \includegraphics[scale = 0.25]{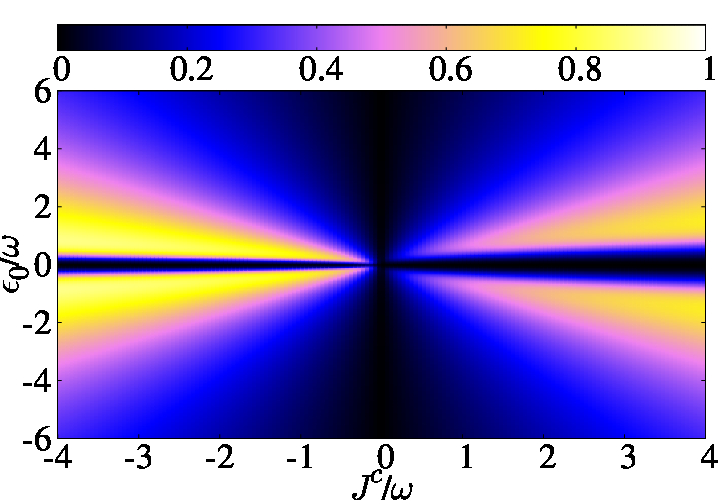}
	    \caption {$A/\omega=0$}
	    \label{fig:9a}
        \end{subfigure}
        \begin{subfigure}[ht!]{0.4\textwidth} 
	    	    \includegraphics[scale = 0.25]{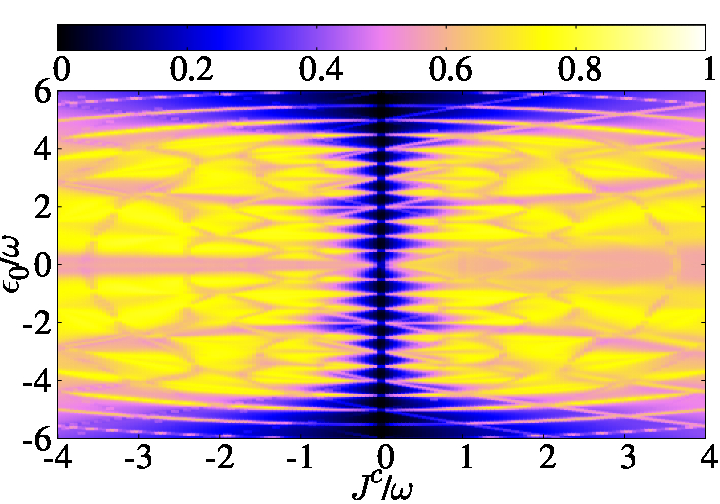}
	    \caption {$A/\omega=3.8$}
	    \label{fig:9b}
        \end{subfigure}
        \caption{Colour map of  $\overline{\overline{C}}$ versus $J^{c}/\omega$ and $\epsilon_{0}/\omega$ for  the capacitive coupling $p=0$, see Hamiltonian Eq.(\ref{eq:9}).The driving amplitudes are $A/\omega=0$ (a) and $A/\omega=3.8$ (b). The initial condition is $|E^{\prime}_{0}\rangle$.(see text for details.}
\label{fig:9}
\end{figure*} 
In Fig.\ref{fig:8a} we show $\overline{\overline{C}}$  without driving for a fixed value  $J^{c}/\omega=-3$. For  $p=0$ (red line) the ground state is separable for $\epsilon_{0}=0$, corresponding to the singlet state $|s_{1}\rangle|s_{2}\rangle$, with $|s_{i}\rangle=(|0\rangle_{i} + |1\rangle_{i})/\sqrt{2}$. On the other hand, for $\epsilon_{0}\neq0$ the ground state is entangled and thus the concurrence increases as the states become mixed. However  for larger values of $\epsilon_{0}$ the state goes asymptotically to a separable state (see Fig.\ref{fig:6b}).

For the $p=1$ (green line)  the ground state is maximally entangled for $\epsilon_{0}=0$, but the concurrence decreases as  $\epsilon_{0}$ increases, analogously to the  $p=0$ case. 
For completeness we include  the coupling $p=0.5$  already studied (black line).

Figure\ref{fig:8c} shows  $\overline{\overline{C}}$  when the driving is on
 for an amplitude of $A/\omega = 3.8$ and $J^{c}/\omega = -3$. For $p=0$  the entanglement  is created in a wide range of $\epsilon_{0}/\omega$, specially near of $\epsilon_{0}= 0$ where the initial state is a separable state (corresponding to a singlet state). For $p=1$ (green line) the entanglement is reduced  near  $\epsilon_{0}=0$.  For larger values of $\epsilon_{0}/\omega$  we observe in both cases a quite similar pattern, with the creation of entanglement due to the driving, but  with  wider resonances compared to $p=0.5$ case.
This is in agreement  with the landscape of avoided crossings in the spectrum of quasienergies, due to the non commutation of the static Hamiltonian with the driving field.

The wider resonances generate a region where  entanglement is quite robust to changes  in the flux detuning. This could be  a tool  to  stabilize the entanglement created by  the driving field. 
In this way, it is interesting to extend the  analysis  studying the dependence of  $\overline{\overline{C}}$ on $p$ and $\epsilon_{0}/\omega$.
Fig. \ref{fig:8b} shows the results without driving for $J^{c}/\omega=-3$.
Two well defined regions can be observed  where there is a qualitative change in the  behaviour of the  concurrence. For $0<p<0.5$ the coupling $J^{c}/2\sigma_{x}^{(1)}\otimes\sigma_{x}^{(2)}$ is the dominant term. In this case, for $\epsilon_{0}\sim 0$ the ground state is separable corresponding to the singlet states $|s_{1}\rangle|s_{2}\rangle$, but as $p$ increases it also does the term $J^{c}/2\sigma_{y}^{(1)}\otimes\sigma_{y}^{(2)}$, and the ground state becomes entangled. 
For the region $0.5<p<1$ the dominant term is $J^{c}
/2\sigma_{y}^{(1)}\otimes\sigma_{y}^{(2)}$ and in this case the ground state remains maximally entangled near $\epsilon_{0}\sim0$ as $p$ decreases.
In Fig.\ref{fig:8d} we present the results for  driving amplitude $A/\omega=3.8$, where important creation of entanglement  with a  rich (and non trivial) pattern of wide  resonances is clearly observed
in an ample range of $p\neq 0.5$.

From the previous analysis it should be clear that although for the symmetric case ($p=0.5$) the resonances are  well defined and sharp, it could be interesting to profit  from   wider resonances created for other values of $p$, in order to control the entanglement induced by the driving field.
 
Besides changing the dominant interaction in the coupling Hamiltonian, one can analize
 the sensitivity of the concurrence  with the coupling strength for a fix value of $p$.
As an example we focus on the  $p=0$ case.
The pattern displayed in  Fig. \ref{fig:9a} in the absence of driving  is  
 consistent with  the  resonant conditions  obtained previously for $(\Delta_{i}, A)\rightarrow 0$.
 In particular notice the regions where the quasi linear behaviour with $J^{c}/\omega$  dominates for  $\epsilon_{0}/\omega \thicksim 0$, turning into  parabolic ones in the plane $(\epsilon_{0}/\omega,J^{c}/\omega)$, for larger values of the flux detuning.
 The concurrence takes different values along  these regions, even for a fixed $J^{c}/\omega$.
The  driving induces a drastic change in this behaviour.
Unlike the longitudinal coupling or the  symmetric transverse coupling case, where the behaviour with $J^{c}/\omega$ for finite driving was quite predictable, we here observe non trivial features. Among others,  is the generation of   an important amount of ${\it homogeneous}$ entanglement for a wide range of   flux detuning and static coupling strength.

Additionally the concurrence exhibits a rather symmetric pattern for $J^{c}/\omega\lessgtr 0$,   similarly to the  \textit{symmetric transverse coupling} case.

\section{Conclusions}
\label{conc}

In this work we have shown that entanglement can be manipulated by external periodic driving fields.
In particular we presented extensive  numerical and analytical results for the  concurrence of a system 
composed by two coupled flux qubits driven by an external ac magnetic flux. 

The main result is that when the system is tuned at or near a multiphoton resonance full control of entanglement is possible: (a) when  the initial state is disentangled one can drive it towards a highly entangled state, and (b) when the initial state is entangled one can strongly reduce entanglement with the driving. 

In the special cases when the driving term in the Hamiltonian and the interaction term in the Hamiltonian commute (longitudinal coupling and symmetric $p=0.5$ transverse coupling), the entanglement control region is within a narrow range of the multiphoton resonances. This shows, for example, as well-defined lines of ``entanglement resonance '' in the $\left\lbrace \epsilon_0,J^{z}(J^{c}) \right\rbrace$ plane in Fig.\ref{fig:4b} and Fig.\ref{fig:7}. One advantage of this case is that once the resonance is exactly tuned, the control of entanglement is almost complete.

In the more common case where the driving Hamiltonian and the interaction Hamiltonian do not commute (as shown here for the $p=0$ transverse coupling), the multiphoton resonances, as seen in terms of an entanglement measure, 
are wide and nearly overlap. In the $\left\lbrace\epsilon_0,J^{c}\right\rbrace$ plane in Fig.\ref{fig:9b} this shows as broad regions where entanglement
can be enhanced starting from a disentangled initial condition. In this case, the control of entanglement is more
robust in parameter space without the need to fine tune at a resonance.  
This suggests that fabricating 
flux qubits that interact through a capacitive (transverse) coupling will be more convenient for
entanglement control through ac driving.

\section*{Acknowlegments}
We acknowledge support from CNEA, CONICET, UNCuyo (P 06/C455) and ANPCyT (PICT2014-1382).

\appendix
\section{\label{ap:A}Concurrence in Floquet basis}
As it was presented in section \ref{sec:II}, the hamiltonian \ref{eq:1} driven by the microwave field, $\epsilon_{i}(t)=\epsilon_{i}+A_{i}\cos(\omega t)$, is periodic in time $\hat{H}(t)=\hat{H}(t+T)$, $T=2\pi/\omega$. So we shall work under the Floquet formalism \cite{shirley}.

Given the global wave function of the system $|\Psi(t)\rangle$, its evolution is governed by the Shr\"odinger equation
\begin{equation}
  \hat{H}(t)|\Psi (t)\rangle =  i\hbar\frac{\partial}{\partial t}|\Psi (t)\rangle.
\label{eq:A1}
\end{equation} 

We can spanned $|\Psi (t)\rangle$ in the Floquet basis
\begin{equation}
  |\Psi (t)\rangle = \sum_{\alpha} a_{\alpha}(t_{0}) e^{-i\gamma_{\alpha}t/\hbar}|u_{\alpha} (t)\rangle,
\label{eq:A2}
\end{equation} where $|u_{\alpha}(t)\rangle=|u_{\alpha}(t+T)\rangle$ is the Floquet state with its respective quasienergy $\gamma_{\alpha}$, $\alpha$ the system index, $a_{\alpha}(t_{0})=\langle u_{\alpha}(t_{0})|\Psi(t_{0})\rangle$  and $|\Psi(t_{0})\rangle$ the initial condition.  If we replace the Eq. \ref{eq:A2} into \ref{eq:A1} we obtain
\begin{equation}
  (H(t) -i\hbar \frac{\partial}{\partial t})|u_{\alpha} (t)\rangle =  \gamma_{\alpha}|u_{\alpha} (t)\rangle
\label{eq:A3}
\end{equation} Solving this equation we obtain the evolution of $|u_{\alpha} (t)\rangle$. It is straightforward to see that the shift $\gamma_{\alpha}\rightarrow\gamma_{\alpha}+n\omega$, $n\in\mathbb{Z}$, leaves unchaged the Eq. \ref{eq:A3} since the quasienergies $\gamma_{\alpha}$ and $\gamma_{\alpha}+n\omega$ correspond to the Floquet states $|u_{\alpha}(t)\rangle$ and $e^{-in\omega t}|u_{\alpha}(t)\rangle$, respectively. 

Now we can calculate the concurrence $C(t,t_{0})=|\langle \Psi(t)|^{*}|\sigma_{y}^{(1)}\otimes \sigma_{y}^{(2)}|\Psi(t)\rangle$, presented in Sec. \ref{sec:II}. Using the expantion $|\Psi(t)\rangle$ in the Floquet basis, we obtain
\begin{equation}
  C(t,t_{0})=|\sum_{\alpha\beta} a_{\alpha}(t_{0})a_{\beta}(t_{0})e^{-i(\gamma_{\alpha}+ \gamma_{\beta})(t-t_{0})} \tilde{C}_{\alpha \beta}(t)|, 
  \label{eq:A4}
\end{equation} with $\gamma_{^{\alpha}_{\beta}}$ the quasienergies, $a_{^{\alpha}_{\beta}}(t_{0})= \langle u_{^{\alpha}_{\beta}}(t_{0})|\Psi(t_{0})\rangle$, $|\Psi(t_{0})\rangle$ the initial condition and $\tilde{C}_{\alpha \beta}(t)= \langle u_{\alpha}(t)|^{*}\sigma_{y}\otimes\sigma_{y} |u_{\beta}(t)\rangle$. We shall take $\hbar=1$.

Using the extended Fourier basis $|u_{\alpha}(t)\rangle= \sum_{k} e^{-i k\omega t}|u_{\alpha}(k)\rangle$ and $\langle u_{\alpha}(t_{0})|= \sum_{q} e^{i q\omega t}\langle u_{\alpha}(q)|$, the Eq. \ref{eq:A4} remains like
\begin{equation}
  C(t,t_{0})= \lvert\sum_{\alpha\beta kk'qq'}\tilde{C}_{\alpha\beta}(k,k')f_{\alpha\beta}(q,q')e^{-i\varphi_{\alpha\beta}^{kk'qq'}(t,t_{0})}\arrowvert,
  \label{eq:A5}
\end{equation} where $\varphi_{\alpha\beta}^{kk'qq'}(t,t_{0})= (\gamma_{\beta}+\gamma_{\alpha} -(k'+k)\omega)t-(\gamma_{\beta}+\gamma_{\alpha} -(q'+q)\omega)t_{0}$ with $\gamma_{^{\alpha}_{\beta}}$ the quasienergies,  $\tilde{C}_{\alpha\beta}(k,k')= \langle u_{\alpha}(k)|^{*} \sigma_{y} \otimes \sigma_{y} |u_{\beta} (k') \rangle$ and $f_{\alpha\beta}(q,q')= a_{\alpha}(q)a_{\beta}(q')$, with $a_{^{\alpha}_{\beta}}(^{q}_{q'})=\langle u_{^{\alpha}_{\beta}}(^{q}_{q'}) | \Psi(t_{0}) \rangle$.

\section{\label{ap:B}Lower bound for  time-averaged concurrence}
Here we calculate a lower bound for the average concurrence $\overline{\overline{C}}$. First we expand the Eq. \ref{eq:A5} as
\begin{equation}
\begin{aligned}
  C(t,t_{0})= \lvert\sum_{\alpha\beta kk'qq'}\tilde{C}_{\alpha\beta}(k,k')f_{\alpha\beta}(q,q')\cos(\varphi_{\alpha\beta}^{kk'qq'}(t,t_{0})\\
  + i \sum_{\alpha\beta kk'qq'}\tilde{C}_{\alpha\beta}(k,k')f_{\alpha\beta}(q,q')\sin(\varphi_{\alpha\beta}^{kk'qq'}(t,t_{0}))\arrowvert
\end{aligned}
  \label{eq:B1}
\end{equation} where we separe it in the real ($Re$) and imaginarie ($Im$) part. Second, using the relations $|z|\geq |Re(z)|,|Im(z)|$, for a complex number $z\in\mathbb{C}$, the Eq. \ref{eq:B1} is changed to
\begin{equation}
\begin{aligned}
  C(t,t_{0})&\geq \lvert\sum_{\alpha\beta kk'qq'}\tilde{C}_{\alpha\beta}(k,k')f_{\alpha\beta}(q,q')\cos(\varphi_{\alpha\beta}^{kk'qq'}(t,t_{0})\arrowvert,\\
  C(t,t_{0})&\geq \lvert\sum_{\alpha\beta kk'qq'}\tilde{C}_{\alpha\beta}(k,k')f_{\alpha\beta}(q,q')\sin(\varphi_{\alpha\beta}^{kk'qq'}(t,t_{0}))\arrowvert,
\end{aligned}
  \label{eq:B2}
\end{equation}  

Now we apply the inequality $\int_{a}^{b}|f(x)|dx \geq |\int_{a}^{b}f(x)dx|$, with $f(x):\mathbb{R}\rightarrow\mathbb{R}$ and $a < b$, to the Eq. \ref{eq:B2}. Here we are inferiorly limiting $\overline{\overline{C}}=\frac{1}{T}\int_{0}^{T}dt\frac{1}{T}\int_{0}^{T}dt_{0} C(t,t_{0})$. Then we obtain two lower bounds $\overline{\overline{C}}\geq C_{I},C_{II}$ such as
\begin{equation}
\begin{aligned}
  C_{I}&=\lvert\sum_{\alpha\beta kk'qq'}\tilde{C}_{\alpha\beta}(k,k')f_{\alpha\beta}(q,q')\overline{\overline{\cos(\varphi_{\alpha\beta}^{kk'qq'}(t,t_{0})}}\arrowvert,\\
  C_{II}&=\lvert\sum_{\alpha\beta kk'qq'}\tilde{C}_{\alpha\beta}(k,k')f_{\alpha\beta}(q,q')\overline{\overline{\sin(\varphi_{\alpha\beta}^{kk'qq'}(t,t_{0}))}}\arrowvert.
\end{aligned}
  \label{eq:B3}
\end{equation} 

Using the trigonometric identities $\sin(A-B)=\sin(A)\cos(B) - \cos(A)\sin(B)$ and $\cos(A-B)=\cos(A)\cos(B) + \sin(A)\sin(B)$, and taking the average over $t,t_{0}$ we obtain
\begin{equation}
\begin{aligned}
  \overline{\overline{\cos(\varphi_{\alpha\beta}^{kk'qq'}(t,t_{0}))}}&= \delta_{\gamma_{\alpha}+\gamma_{\beta}-n\omega,0}\,\delta_{\gamma_{\alpha}+\gamma_{\beta}-m\omega,0} \\
  &+ \delta_{\gamma_{\alpha}+\gamma_{\beta}-n\omega,\pi/(2T)}\,\delta_{\gamma_{\alpha}+\gamma_{\beta}-m\omega,\pi/(2T)},\\
  \overline{\overline{\sin(\varphi_{\alpha\beta}^{kk'qq'}(t,t_{0}))}}&= \delta_{\gamma_{\alpha}+\gamma_{\beta}-n\omega,\pi/(2T)}\,\delta_{\gamma_{\alpha}+\gamma_{\beta}-m\omega,0} \\
  &- \delta_{\gamma_{\alpha}+\gamma_{\beta}-n\omega,0}\,\delta_{\gamma_{\alpha}+\gamma_{\beta}-m\omega,\pi/(2T)},
\end{aligned}
  \label{eq:B4}
\end{equation} with $n=k+k'$ and $m=q+q'$. Replacing $T=2\pi/\omega$ and ordening the terms, we obtain
\begin{equation}
\begin{aligned}
  \overline{\overline{\cos(\varphi_{\alpha\beta}^{kk'qq'}(t,t_{0}))}}&= \delta_{\gamma_{\alpha}+\gamma_{\beta},n\omega}\,\delta_{\gamma_{\alpha}+\gamma_{\beta},m\omega} \\
  &+ \delta_{\gamma_{\alpha}+\gamma_{\beta},(n+1/4)\omega}\,\delta_{\gamma_{\alpha}+\gamma_{\beta},(m+1/4)\omega},\\
  \overline{\overline{\sin(\varphi_{\alpha\beta}^{kk'qq'}(t,t_{0}))}}&= \delta_{\gamma_{\alpha}+\gamma_{\beta},(n+1/4)\omega}\,\delta_{\gamma_{\alpha}+\gamma_{\beta},m\omega} \\
  &- \delta_{\gamma_{\alpha}+\gamma_{\beta},n\omega}\,\delta_{\gamma_{\alpha}+\gamma_{\beta},(m+1/4)\omega}.
\end{aligned}
  \label{eq:B5}
\end{equation} 

We suppose the main contribution to concurrence is near the resonances condition $\gamma_{\alpha}+\gamma_{\beta}=n\omega$, this is equivalent to a rotating wave approximation disregarding fast oscillating terms. The averages over $t,t_{0}$ remain 
\begin{equation}
\begin{aligned}
  \overline{\overline{\cos(\varphi_{\alpha\beta}^{kk'qq'}(t,t_{0}))}}&\sim \delta_{\gamma_{\alpha}+\gamma_{\beta},n\omega}\,\delta_{\gamma_{\alpha}+\gamma_{\beta},m\omega},\\
  \overline{\overline{\sin(\varphi_{\alpha\beta}^{kk'qq'}(t,t_{0}))}}&\sim0.
\end{aligned}
  \label{eq:B6}
\end{equation} 

Using the last result we obtain $\overline{\overline{C}} \geq C_{I} \geq C_{II}=0$, then the corresponding lower bound expression is
\begin{equation}
\begin{aligned}
  C_{I}\sim\lvert\sum_{^{\alpha\beta}_{knqm}}\tilde{C}_{\alpha\beta}(k,n-k)f_{\alpha\beta}(q,m-q) \delta_{\gamma_{\alpha}+\gamma_{\beta},n\omega}\,\delta_{\gamma_{\alpha}+\gamma_{\beta},m\omega}\arrowvert,\\
\end{aligned}
  \label{eq:B7}
\end{equation} 

Given the Eq.\ref{eq:B7}, it is necessary evaluate the condition $\gamma_{\beta}=-\gamma_{\alpha}+n\omega$ (also $m$) on the involved Floquet states. We know that for a given Floquet state $|u_{\beta}(t)\rangle$ its evolution is governed by $e^{-i\gamma_{\beta}}|u_{\beta}(t)\rangle$. Therefore considering the resonance condition we obtain
\begin{equation}
\begin{aligned}
  e^{-i\gamma_{\beta}}|u_{\beta}(t)\rangle&=e^{i\gamma_{\alpha}}e^{-in\omega}|u_{\beta}(t)\rangle,\\
  &=e^{i\gamma_{\alpha}}e^{-in\omega}\sum_{k}e^{ik\omega t}|u_{\beta}(k)\rangle\\
  &=e^{i\gamma_{\alpha}}\sum_{k}e^{i(k-n)\omega t}|u_{\beta}(k)\rangle\\
  &=e^{i\gamma_{\alpha}}\sum_{k}e^{ik\omega t}|u_{\beta}(k-n)\rangle\\
  &=e^{i\gamma_{\alpha}}\sum_{k}e^{-ik\omega t}|u_{\beta}(n-k)\rangle.
\end{aligned}
  \label{eq:B8}
\end{equation}  For another part, we can compare the last result to  $(e^{-i\gamma_{\alpha}}|u_{\alpha}(t)\rangle)^{*}$, where
\begin{equation}
\begin{aligned}
  (e^{-i\gamma_{\alpha}}|u_{\alpha}(t)\rangle)^{*}&=e^{i\gamma_{\alpha}}|u_{\alpha}(t)\rangle^{*},\\
  &=e^{i\gamma_{\alpha}}\sum_{k}e^{-ik\omega t}|u_{\alpha}(k)\rangle^{*}\\
  &=e^{i\gamma_{\alpha}}\sum_{k}e^{-ik\omega t}|u_{\alpha}(-k)\rangle.
\end{aligned}
  \label{eq:B9}
\end{equation} Then, from Eq.\ref{eq:B8} and \ref{eq:B9}, we obtain an equivalence relation:
\begin{equation}
  \gamma_{\beta} \rightarrow -\gamma_{\alpha} + n\omega \Rightarrow |u_{\beta}(n-k)\rangle \rightarrow |u_{\alpha}(-k)\rangle. 
  \label{eq:B10}
\end{equation}

Using the result \ref{eq:B10} with $n=k+k'$ and $m=q+q'$, the lower bound in Eq. \ref{eq:B7} remains like
\begin{equation}
\begin{aligned}
  C_{I} &\sim |\sum_{\alpha kq}\tilde{C}_{\alpha\alpha}(k,-k)f_{\alpha\alpha}(q,-q)|,
\end{aligned}
  \label{eq:B11}
\end{equation} where  $\tilde{C}_{\alpha\alpha}(k,-k)= \langle u_{\alpha}(k)|^{*} \sigma_{y} \otimes \sigma_{y} |u_{\alpha} (-k) \rangle$ and $f_{\alpha\alpha}(q,-q)= a_{\alpha}(q)a_{\alpha}(-q)$, with $a_{\alpha}(q)=\langle u_{\alpha}(q) | \Psi(t_{0}) \rangle$. 

From Eq.\ref{eq:B11}, we identify  a contribution of the form
\begin{equation}
\begin{aligned}
  \sum_{k}\tilde{C}_{\alpha\beta}(k,-k)&=\frac{1}{T}\int_{0}^{T}dt\,\tilde{C}_{\alpha\alpha}(t)=\overline{\tilde{C}_{\alpha\alpha}(t)},
\end{aligned}
  \label{eq:B12}
\end{equation} where $\tilde{C}_{\alpha\alpha}(t)=\langle u_{\alpha}(t)|^{*} \sigma_{y} \otimes \sigma_{y} |u_{\alpha} (t) \rangle$ is the \textit{Floquet preconcurrence}. Also, using the property $|u_{\alpha}(-q)\rangle = |u_{\alpha}(q)\rangle^{*}$ we obtain 
\begin{equation}
\begin{aligned}
  f_{\alpha\alpha}(-q,q)= a_{\alpha}(-q)a_{\alpha}(q)=a^{*}_{\alpha}(q)a_{\alpha}(q)=|a_{\alpha}(q)|^{2},
\end{aligned}
  \label{eq:B13}
\end{equation} corresponding to the amplitude of the Floquet states projections over the initial condition. 

Finally remplacing equations \ref{eq:B12} and \ref{eq:B13} in \ref{eq:B11} we obtain 
\begin{equation}
\begin{aligned}
  C_{I} &\sim |\sum_{\alpha}\overline{\tilde{C}_{\alpha\alpha}(t)}\sum_{q}|a_{\alpha}(q)|^{2}|,
\end{aligned}
  \label{eq:B14}
\end{equation} This expression is useful for a non-entangled initial state since it represents the minimal entanglement creation.

\bibliography{references1}

\begin{thebibliography}{35}%
\makeatletter
\providecommand \@ifxundefined [1]{%
 \@ifx{#1\undefined}
}%
\providecommand \@ifnum [1]{%
 \ifnum #1\expandafter \@firstoftwo
 \else \expandafter \@secondoftwo
 \fi
}%
\providecommand \@ifx [1]{%
 \ifx #1\expandafter \@firstoftwo
 \else \expandafter \@secondoftwo
 \fi
}%
\providecommand \natexlab [1]{#1}%
\providecommand \enquote  [1]{``#1''}%
\providecommand \bibnamefont  [1]{#1}%
\providecommand \bibfnamefont [1]{#1}%
\providecommand \citenamefont [1]{#1}%
\providecommand \href@noop [0]{\@secondoftwo}%
\providecommand \href [0]{\begingroup \@sanitize@url \@href}%
\providecommand \@href[1]{\@@startlink{#1}\@@href}%
\providecommand \@@href[1]{\endgroup#1\@@endlink}%
\providecommand \@sanitize@url [0]{\catcode `\\12\catcode `\$12\catcode
  `\&12\catcode `\#12\catcode `\^12\catcode `\_12\catcode `\%12\relax}%
\providecommand \@@startlink[1]{}%
\providecommand \@@endlink[0]{}%
\providecommand \url  [0]{\begingroup\@sanitize@url \@url }%
\providecommand \@url [1]{\endgroup\@href {#1}{\urlprefix }}%
\providecommand \urlprefix  [0]{URL }%
\providecommand \Eprint [0]{\href }%
\providecommand \doibase [0]{http://dx.doi.org/}%
\providecommand \selectlanguage [0]{\@gobble}%
\providecommand \bibinfo  [0]{\@secondoftwo}%
\providecommand \bibfield  [0]{\@secondoftwo}%
\providecommand \translation [1]{[#1]}%
\providecommand \BibitemOpen [0]{}%
\providecommand \bibitemStop [0]{}%
\providecommand \bibitemNoStop [0]{.\EOS\space}%
\providecommand \EOS [0]{\spacefactor3000\relax}%
\providecommand \BibitemShut  [1]{\csname bibitem#1\endcsname}%
\let\auto@bib@innerbib\@empty
\bibitem [{\citenamefont {Orlando}\ \emph {et~al.}(1999)\citenamefont
  {Orlando}, \citenamefont {Mooij}, \citenamefont {Tian}, \citenamefont
  {van~der Wal}, \citenamefont {Levitov}, \citenamefont {Lloyd},\ and\
  \citenamefont {Mazo}}]{orlando}%
  \BibitemOpen
  \bibfield  {author} {\bibinfo {author} {\bibfnamefont {T.~P.}\ \bibnamefont
  {Orlando}}, \bibinfo {author} {\bibfnamefont {J.~E.}\ \bibnamefont {Mooij}},
  \bibinfo {author} {\bibfnamefont {L.}~\bibnamefont {Tian}}, \bibinfo {author}
  {\bibfnamefont {C.~H.}\ \bibnamefont {van~der Wal}}, \bibinfo {author}
  {\bibfnamefont {L.~S.}\ \bibnamefont {Levitov}}, \bibinfo {author}
  {\bibfnamefont {S.}~\bibnamefont {Lloyd}}, \ and\ \bibinfo {author}
  {\bibfnamefont {J.~J.}\ \bibnamefont {Mazo}},\ }\href {\doibase
  10.1103/PhysRevB.60.15398} {\bibfield  {journal} {\bibinfo  {journal} {Phys.
  Rev. B}\ }\textbf {\bibinfo {volume} {60}},\ \bibinfo {pages} {15398}
  (\bibinfo {year} {1999})}\BibitemShut {NoStop}%
\bibitem [{\citenamefont {Friedman}\ \emph {et~al.}(2000)\citenamefont
  {Friedman}, \citenamefont {Patel}, \citenamefont {Chen}, \citenamefont
  {Tolpygo},\ and\ \citenamefont {Lukens}}]{friedman}%
  \BibitemOpen
  \bibfield  {author} {\bibinfo {author} {\bibfnamefont {J.~R.}\ \bibnamefont
  {Friedman}}, \bibinfo {author} {\bibfnamefont {V.}~\bibnamefont {Patel}},
  \bibinfo {author} {\bibfnamefont {W.}~\bibnamefont {Chen}}, \bibinfo {author}
  {\bibfnamefont {S.~K.}\ \bibnamefont {Tolpygo}}, \ and\ \bibinfo {author}
  {\bibfnamefont {J.~E.}\ \bibnamefont {Lukens}},\ }\href
  {http://dx.doi.org/10.1038/35017505} {\bibfield  {journal} {\bibinfo
  {journal} {Nature}\ }\textbf {\bibinfo {volume} {406}},\ \bibinfo {pages}
  {43} (\bibinfo {year} {2000})}\BibitemShut {NoStop}%
\bibitem [{\citenamefont {van~der Wal}\ \emph {et~al.}(2000)\citenamefont
  {van~der Wal}, \citenamefont {ter Haar}, \citenamefont {Wilhelm},
  \citenamefont {Schouten}, \citenamefont {Harmans}, \citenamefont {Orlando},
  \citenamefont {Lloyd},\ and\ \citenamefont {Mooij}}]{vanderwal}%
  \BibitemOpen
  \bibfield  {author} {\bibinfo {author} {\bibfnamefont {C.~H.}\ \bibnamefont
  {van~der Wal}}, \bibinfo {author} {\bibfnamefont {A.~C.~J.}\ \bibnamefont
  {ter Haar}}, \bibinfo {author} {\bibfnamefont {F.~K.}\ \bibnamefont
  {Wilhelm}}, \bibinfo {author} {\bibfnamefont {R.~N.}\ \bibnamefont
  {Schouten}}, \bibinfo {author} {\bibfnamefont {C.~J. P.~M.}\ \bibnamefont
  {Harmans}}, \bibinfo {author} {\bibfnamefont {T.~P.}\ \bibnamefont
  {Orlando}}, \bibinfo {author} {\bibfnamefont {S.}~\bibnamefont {Lloyd}}, \
  and\ \bibinfo {author} {\bibfnamefont {J.~E.}\ \bibnamefont {Mooij}},\ }\href
  {\doibase 10.1126/science.290.5492.773} {\bibfield  {journal} {\bibinfo
  {journal} {Science}\ }\textbf {\bibinfo {volume} {290}},\ \bibinfo {pages}
  {773} (\bibinfo {year} {2000})},\ \Eprint
  {http://arxiv.org/abs/http://science.sciencemag.org/content/290/5492/773.full.pdf}
  {http://science.sciencemag.org/content/290/5492/773.full.pdf} \BibitemShut
  {NoStop}%
\bibitem [{\citenamefont {Martinis}\ \emph {et~al.}(2002)\citenamefont
  {Martinis}, \citenamefont {Nam}, \citenamefont {Aumentado},\ and\
  \citenamefont {Urbina}}]{martinis}%
  \BibitemOpen
  \bibfield  {author} {\bibinfo {author} {\bibfnamefont {J.~M.}\ \bibnamefont
  {Martinis}}, \bibinfo {author} {\bibfnamefont {S.}~\bibnamefont {Nam}},
  \bibinfo {author} {\bibfnamefont {J.}~\bibnamefont {Aumentado}}, \ and\
  \bibinfo {author} {\bibfnamefont {C.}~\bibnamefont {Urbina}},\ }\href
  {\doibase 10.1103/PhysRevLett.89.117901} {\bibfield  {journal} {\bibinfo
  {journal} {Phys. Rev. Lett.}\ }\textbf {\bibinfo {volume} {89}},\ \bibinfo
  {pages} {117901} (\bibinfo {year} {2002})}\BibitemShut {NoStop}%
\bibitem [{\citenamefont {Oliver}\ \emph {et~al.}(2005)\citenamefont {Oliver},
  \citenamefont {Yu}, \citenamefont {Lee}, \citenamefont {Berggren},
  \citenamefont {Levitov},\ and\ \citenamefont {Orlando}}]{oliver}%
  \BibitemOpen
  \bibfield  {author} {\bibinfo {author} {\bibfnamefont {W.~D.}\ \bibnamefont
  {Oliver}}, \bibinfo {author} {\bibfnamefont {Y.}~\bibnamefont {Yu}}, \bibinfo
  {author} {\bibfnamefont {J.~C.}\ \bibnamefont {Lee}}, \bibinfo {author}
  {\bibfnamefont {K.~K.}\ \bibnamefont {Berggren}}, \bibinfo {author}
  {\bibfnamefont {L.~S.}\ \bibnamefont {Levitov}}, \ and\ \bibinfo {author}
  {\bibfnamefont {T.~P.}\ \bibnamefont {Orlando}},\ }\href {\doibase
  10.1126/science.1119678} {\bibfield  {journal} {\bibinfo  {journal}
  {Science}\ }\textbf {\bibinfo {volume} {310}},\ \bibinfo {pages} {1653}
  (\bibinfo {year} {2005})},\ \Eprint
  {http://arxiv.org/abs/http://science.sciencemag.org/content/310/5754/1653.full.pdf}
  {http://science.sciencemag.org/content/310/5754/1653.full.pdf} \BibitemShut
  {NoStop}%
\bibitem [{\citenamefont {Majer}\ \emph {et~al.}(2005)\citenamefont {Majer},
  \citenamefont {Paauw}, \citenamefont {ter Haar}, \citenamefont {Harmans},\
  and\ \citenamefont {Mooij}}]{majer}%
  \BibitemOpen
  \bibfield  {author} {\bibinfo {author} {\bibfnamefont {J.~B.}\ \bibnamefont
  {Majer}}, \bibinfo {author} {\bibfnamefont {F.~G.}\ \bibnamefont {Paauw}},
  \bibinfo {author} {\bibfnamefont {A.~C.~J.}\ \bibnamefont {ter Haar}},
  \bibinfo {author} {\bibfnamefont {C.~J. P.~M.}\ \bibnamefont {Harmans}}, \
  and\ \bibinfo {author} {\bibfnamefont {J.~E.}\ \bibnamefont {Mooij}},\ }\href
  {\doibase 10.1103/PhysRevLett.94.090501} {\bibfield  {journal} {\bibinfo
  {journal} {Phys. Rev. Lett.}\ }\textbf {\bibinfo {volume} {94}},\ \bibinfo
  {pages} {090501} (\bibinfo {year} {2005})}\BibitemShut {NoStop}%
\bibitem [{\citenamefont {Strauch}\ \emph {et~al.}(2003)\citenamefont
  {Strauch}, \citenamefont {Johnson}, \citenamefont {Dragt}, \citenamefont
  {Lobb}, \citenamefont {Anderson},\ and\ \citenamefont {Wellstood}}]{strauch}%
  \BibitemOpen
  \bibfield  {author} {\bibinfo {author} {\bibfnamefont {F.~W.}\ \bibnamefont
  {Strauch}}, \bibinfo {author} {\bibfnamefont {P.~R.}\ \bibnamefont
  {Johnson}}, \bibinfo {author} {\bibfnamefont {A.~J.}\ \bibnamefont {Dragt}},
  \bibinfo {author} {\bibfnamefont {C.~J.}\ \bibnamefont {Lobb}}, \bibinfo
  {author} {\bibfnamefont {J.~R.}\ \bibnamefont {Anderson}}, \ and\ \bibinfo
  {author} {\bibfnamefont {F.~C.}\ \bibnamefont {Wellstood}},\ }\href {\doibase
  10.1103/PhysRevLett.91.167005} {\bibfield  {journal} {\bibinfo  {journal}
  {Phys. Rev. Lett.}\ }\textbf {\bibinfo {volume} {91}},\ \bibinfo {pages}
  {167005} (\bibinfo {year} {2003})}\BibitemShut {NoStop}%
\bibitem [{\citenamefont {Izmalkov}\ \emph {et~al.}(2004)\citenamefont
  {Izmalkov}, \citenamefont {Grajcar}, \citenamefont {Il'ichev}, \citenamefont
  {Wagner}, \citenamefont {Meyer}, \citenamefont {Smirnov}, \citenamefont
  {Amin}, \citenamefont {van~den Brink},\ and\ \citenamefont
  {Zagoskin}}]{izmalkov}%
  \BibitemOpen
  \bibfield  {author} {\bibinfo {author} {\bibfnamefont {A.}~\bibnamefont
  {Izmalkov}}, \bibinfo {author} {\bibfnamefont {M.}~\bibnamefont {Grajcar}},
  \bibinfo {author} {\bibfnamefont {E.}~\bibnamefont {Il'ichev}}, \bibinfo
  {author} {\bibfnamefont {T.}~\bibnamefont {Wagner}}, \bibinfo {author}
  {\bibfnamefont {H.-G.}\ \bibnamefont {Meyer}}, \bibinfo {author}
  {\bibfnamefont {A.~Y.}\ \bibnamefont {Smirnov}}, \bibinfo {author}
  {\bibfnamefont {M.~H.~S.}\ \bibnamefont {Amin}}, \bibinfo {author}
  {\bibfnamefont {A.~M.}\ \bibnamefont {van~den Brink}}, \ and\ \bibinfo
  {author} {\bibfnamefont {A.~M.}\ \bibnamefont {Zagoskin}},\ }\href {\doibase
  10.1103/PhysRevLett.93.037003} {\bibfield  {journal} {\bibinfo  {journal}
  {Phys. Rev. Lett.}\ }\textbf {\bibinfo {volume} {93}},\ \bibinfo {pages}
  {037003} (\bibinfo {year} {2004})}\BibitemShut {NoStop}%
\bibitem [{\citenamefont {Grajcar}\ \emph {et~al.}(2005)\citenamefont
  {Grajcar}, \citenamefont {Izmalkov}, \citenamefont {van~der Ploeg},
  \citenamefont {Linzen}, \citenamefont {Il'ichev}, \citenamefont {Wagner},
  \citenamefont {H\"ubner}, \citenamefont {Meyer}, \citenamefont {Maassen
  van~den Brink}, \citenamefont {Uchaikin},\ and\ \citenamefont
  {Zagoskin}}]{grajcar}%
  \BibitemOpen
  \bibfield  {author} {\bibinfo {author} {\bibfnamefont {M.}~\bibnamefont
  {Grajcar}}, \bibinfo {author} {\bibfnamefont {A.}~\bibnamefont {Izmalkov}},
  \bibinfo {author} {\bibfnamefont {S.~H.~W.}\ \bibnamefont {van~der Ploeg}},
  \bibinfo {author} {\bibfnamefont {S.}~\bibnamefont {Linzen}}, \bibinfo
  {author} {\bibfnamefont {E.}~\bibnamefont {Il'ichev}}, \bibinfo {author}
  {\bibfnamefont {T.}~\bibnamefont {Wagner}}, \bibinfo {author} {\bibfnamefont
  {U.}~\bibnamefont {H\"ubner}}, \bibinfo {author} {\bibfnamefont {H.-G.}\
  \bibnamefont {Meyer}}, \bibinfo {author} {\bibfnamefont {A.}~\bibnamefont
  {Maassen van~den Brink}}, \bibinfo {author} {\bibfnamefont {S.}~\bibnamefont
  {Uchaikin}}, \ and\ \bibinfo {author} {\bibfnamefont {A.~M.}\ \bibnamefont
  {Zagoskin}},\ }\href {\doibase 10.1103/PhysRevB.72.020503} {\bibfield
  {journal} {\bibinfo  {journal} {Phys. Rev. B}\ }\textbf {\bibinfo {volume}
  {72}},\ \bibinfo {pages} {020503} (\bibinfo {year} {2005})}\BibitemShut
  {NoStop}%
\bibitem [{\citenamefont {Yamamoto}\ \emph {et~al.}(2003)\citenamefont
  {Yamamoto}, \citenamefont {Pashkin}, \citenamefont {Astafiev}, \citenamefont
  {Nakamura},\ and\ \citenamefont {Tsai}}]{yamamoto}%
  \BibitemOpen
  \bibfield  {author} {\bibinfo {author} {\bibfnamefont {T.}~\bibnamefont
  {Yamamoto}}, \bibinfo {author} {\bibfnamefont {Y.~A.}\ \bibnamefont
  {Pashkin}}, \bibinfo {author} {\bibfnamefont {O.}~\bibnamefont {Astafiev}},
  \bibinfo {author} {\bibfnamefont {Y.}~\bibnamefont {Nakamura}}, \ and\
  \bibinfo {author} {\bibfnamefont {J.~S.}\ \bibnamefont {Tsai}},\ }\href
  {http://dx.doi.org/10.1038/nature02015} {\bibfield  {journal} {\bibinfo
  {journal} {Nature}\ }\textbf {\bibinfo {volume} {425}},\ \bibinfo {pages}
  {941} (\bibinfo {year} {2003})}\BibitemShut {NoStop}%
\bibitem [{\citenamefont {Mooij}\ \emph {et~al.}(1999)\citenamefont {Mooij},
  \citenamefont {Orlando}, \citenamefont {Levitov}, \citenamefont {Tian},
  \citenamefont {van~der Wal},\ and\ \citenamefont {Lloyd}}]{mooij}%
  \BibitemOpen
  \bibfield  {author} {\bibinfo {author} {\bibfnamefont {J.~E.}\ \bibnamefont
  {Mooij}}, \bibinfo {author} {\bibfnamefont {T.~P.}\ \bibnamefont {Orlando}},
  \bibinfo {author} {\bibfnamefont {L.}~\bibnamefont {Levitov}}, \bibinfo
  {author} {\bibfnamefont {L.}~\bibnamefont {Tian}}, \bibinfo {author}
  {\bibfnamefont {C.~H.}\ \bibnamefont {van~der Wal}}, \ and\ \bibinfo {author}
  {\bibfnamefont {S.}~\bibnamefont {Lloyd}},\ }\href {\doibase
  10.1126/science.285.5430.1036} {\bibfield  {journal} {\bibinfo  {journal}
  {Science}\ }\textbf {\bibinfo {volume} {285}},\ \bibinfo {pages} {1036}
  (\bibinfo {year} {1999})},\ \Eprint
  {http://arxiv.org/abs/http://science.sciencemag.org/content/285/5430/1036.full.pdf}
  {http://science.sciencemag.org/content/285/5430/1036.full.pdf} \BibitemShut
  {NoStop}%
\bibitem [{\citenamefont {van~der Ploeg}\ \emph {et~al.}(2007)\citenamefont
  {van~der Ploeg}, \citenamefont {Izmalkov}, \citenamefont {van~den Brink},
  \citenamefont {H\"ubner}, \citenamefont {Grajcar}, \citenamefont {Il'ichev},
  \citenamefont {Meyer},\ and\ \citenamefont {Zagoskin}}]{nori1}%
  \BibitemOpen
  \bibfield  {author} {\bibinfo {author} {\bibfnamefont {S.~H.~W.}\
  \bibnamefont {van~der Ploeg}}, \bibinfo {author} {\bibfnamefont
  {A.}~\bibnamefont {Izmalkov}}, \bibinfo {author} {\bibfnamefont {A.~M.}\
  \bibnamefont {van~den Brink}}, \bibinfo {author} {\bibfnamefont
  {U.}~\bibnamefont {H\"ubner}}, \bibinfo {author} {\bibfnamefont
  {M.}~\bibnamefont {Grajcar}}, \bibinfo {author} {\bibfnamefont
  {E.}~\bibnamefont {Il'ichev}}, \bibinfo {author} {\bibfnamefont {H.-G.}\
  \bibnamefont {Meyer}}, \ and\ \bibinfo {author} {\bibfnamefont {A.~M.}\
  \bibnamefont {Zagoskin}},\ }\href {\doibase 10.1103/PhysRevLett.98.057004}
  {\bibfield  {journal} {\bibinfo  {journal} {Phys. Rev. Lett.}\ }\textbf
  {\bibinfo {volume} {98}},\ \bibinfo {pages} {057004} (\bibinfo {year}
  {2007})}\BibitemShut {NoStop}%
\bibitem [{\citenamefont {Wendin}\ and\ \citenamefont
  {Shumeiko}(2005)}]{wendin}%
  \BibitemOpen
  \bibfield  {author} {\bibinfo {author} {\bibfnamefont {G.}~\bibnamefont
  {Wendin}}\ and\ \bibinfo {author} {\bibfnamefont {V.}~\bibnamefont
  {Shumeiko}},\ }\href {arXiv:cond-mat/0508729v1} {\  (\bibinfo {year}
  {2005})}\BibitemShut {NoStop}%
\bibitem [{\citenamefont {Plourde}\ \emph {et~al.}(2004)\citenamefont
  {Plourde}, \citenamefont {Zhang}, \citenamefont {Whaley}, \citenamefont
  {Wilhelm}, \citenamefont {Robertson}, \citenamefont {Hime}, \citenamefont
  {Linzen}, \citenamefont {Reichardt}, \citenamefont {Wu},\ and\ \citenamefont
  {Clarke}}]{plourde}%
  \BibitemOpen
  \bibfield  {author} {\bibinfo {author} {\bibfnamefont {B.~L.~T.}\
  \bibnamefont {Plourde}}, \bibinfo {author} {\bibfnamefont {J.}~\bibnamefont
  {Zhang}}, \bibinfo {author} {\bibfnamefont {K.~B.}\ \bibnamefont {Whaley}},
  \bibinfo {author} {\bibfnamefont {F.~K.}\ \bibnamefont {Wilhelm}}, \bibinfo
  {author} {\bibfnamefont {T.~L.}\ \bibnamefont {Robertson}}, \bibinfo {author}
  {\bibfnamefont {T.}~\bibnamefont {Hime}}, \bibinfo {author} {\bibfnamefont
  {S.}~\bibnamefont {Linzen}}, \bibinfo {author} {\bibfnamefont {P.~A.}\
  \bibnamefont {Reichardt}}, \bibinfo {author} {\bibfnamefont {C.-E.}\
  \bibnamefont {Wu}}, \ and\ \bibinfo {author} {\bibfnamefont {J.}~\bibnamefont
  {Clarke}},\ }\href {\doibase 10.1103/PhysRevB.70.140501} {\bibfield
  {journal} {\bibinfo  {journal} {Phys. Rev. B}\ }\textbf {\bibinfo {volume}
  {70}},\ \bibinfo {pages} {140501} (\bibinfo {year} {2004})}\BibitemShut
  {NoStop}%
\bibitem [{\citenamefont {Harrabi}\ \emph {et~al.}(2009)\citenamefont
  {Harrabi}, \citenamefont {Yoshihara}, \citenamefont {Niskanen}, \citenamefont
  {Nakamura},\ and\ \citenamefont {Tsai}}]{harrabi}%
  \BibitemOpen
  \bibfield  {author} {\bibinfo {author} {\bibfnamefont {K.}~\bibnamefont
  {Harrabi}}, \bibinfo {author} {\bibfnamefont {F.}~\bibnamefont {Yoshihara}},
  \bibinfo {author} {\bibfnamefont {A.~O.}\ \bibnamefont {Niskanen}}, \bibinfo
  {author} {\bibfnamefont {Y.}~\bibnamefont {Nakamura}}, \ and\ \bibinfo
  {author} {\bibfnamefont {J.~S.}\ \bibnamefont {Tsai}},\ }\href {\doibase
  10.1103/PhysRevB.79.020507} {\bibfield  {journal} {\bibinfo  {journal} {Phys.
  Rev. B}\ }\textbf {\bibinfo {volume} {79}},\ \bibinfo {pages} {020507}
  (\bibinfo {year} {2009})}\BibitemShut {NoStop}%
\bibitem [{\citenamefont {de~Groot}\ \emph {et~al.}(2010)\citenamefont
  {de~Groot}, \citenamefont {Lisenfeld}, \citenamefont {Schouten},
  \citenamefont {Ashhab}, \citenamefont {Lupascu}, \citenamefont {Harmans},\
  and\ \citenamefont {Mooij}}]{degroot}%
  \BibitemOpen
  \bibfield  {author} {\bibinfo {author} {\bibfnamefont {P.~C.}\ \bibnamefont
  {de~Groot}}, \bibinfo {author} {\bibfnamefont {J.}~\bibnamefont {Lisenfeld}},
  \bibinfo {author} {\bibfnamefont {R.~N.}\ \bibnamefont {Schouten}}, \bibinfo
  {author} {\bibfnamefont {S.}~\bibnamefont {Ashhab}}, \bibinfo {author}
  {\bibfnamefont {A.}~\bibnamefont {Lupascu}}, \bibinfo {author} {\bibfnamefont
  {C.~J. P.~M.}\ \bibnamefont {Harmans}}, \ and\ \bibinfo {author}
  {\bibfnamefont {J.~E.}\ \bibnamefont {Mooij}},\ }\href
  {http://dx.doi.org/10.1038/nphys1733} {\bibfield  {journal} {\bibinfo
  {journal} {Nat Phys}\ }\textbf {\bibinfo {volume} {6}},\ \bibinfo {pages}
  {763} (\bibinfo {year} {2010})}\BibitemShut {NoStop}%
\bibitem [{\citenamefont {Shevchenko}\ \emph {et~al.}(2010)\citenamefont
  {Shevchenko}, \citenamefont {Ashhab},\ and\ \citenamefont
  {Nori}}]{shevchenko}%
  \BibitemOpen
  \bibfield  {author} {\bibinfo {author} {\bibfnamefont {S.}~\bibnamefont
  {Shevchenko}}, \bibinfo {author} {\bibfnamefont {S.}~\bibnamefont {Ashhab}},
  \ and\ \bibinfo {author} {\bibfnamefont {F.}~\bibnamefont {Nori}},\ }\href
  {\doibase http://dx.doi.org/10.1016/j.physrep.2010.03.002} {\bibfield
  {journal} {\bibinfo  {journal} {Physics Reports}\ }\textbf {\bibinfo {volume}
  {492}},\ \bibinfo {pages} {1 } (\bibinfo {year} {2010})}\BibitemShut
  {NoStop}%
\bibitem [{\citenamefont {Berns}\ \emph {et~al.}(2006)\citenamefont {Berns},
  \citenamefont {Oliver}, \citenamefont {Valenzuela}, \citenamefont {Shytov},
  \citenamefont {Berggren}, \citenamefont {Levitov},\ and\ \citenamefont
  {Orlando}}]{berns}%
  \BibitemOpen
  \bibfield  {author} {\bibinfo {author} {\bibfnamefont {D.~M.}\ \bibnamefont
  {Berns}}, \bibinfo {author} {\bibfnamefont {W.~D.}\ \bibnamefont {Oliver}},
  \bibinfo {author} {\bibfnamefont {S.~O.}\ \bibnamefont {Valenzuela}},
  \bibinfo {author} {\bibfnamefont {A.~V.}\ \bibnamefont {Shytov}}, \bibinfo
  {author} {\bibfnamefont {K.~K.}\ \bibnamefont {Berggren}}, \bibinfo {author}
  {\bibfnamefont {L.~S.}\ \bibnamefont {Levitov}}, \ and\ \bibinfo {author}
  {\bibfnamefont {T.~P.}\ \bibnamefont {Orlando}},\ }\href {\doibase
  10.1103/PhysRevLett.97.150502} {\bibfield  {journal} {\bibinfo  {journal}
  {Phys. Rev. Lett.}\ }\textbf {\bibinfo {volume} {97}},\ \bibinfo {pages}
  {150502} (\bibinfo {year} {2006})}\BibitemShut {NoStop}%
\bibitem [{\citenamefont {Rudner}\ \emph {et~al.}(2008)\citenamefont {Rudner},
  \citenamefont {Shytov}, \citenamefont {Levitov}, \citenamefont {Berns},
  \citenamefont {Oliver}, \citenamefont {Valenzuela},\ and\ \citenamefont
  {Orlando}}]{rudner}%
  \BibitemOpen
  \bibfield  {author} {\bibinfo {author} {\bibfnamefont {M.~S.}\ \bibnamefont
  {Rudner}}, \bibinfo {author} {\bibfnamefont {A.~V.}\ \bibnamefont {Shytov}},
  \bibinfo {author} {\bibfnamefont {L.~S.}\ \bibnamefont {Levitov}}, \bibinfo
  {author} {\bibfnamefont {D.~M.}\ \bibnamefont {Berns}}, \bibinfo {author}
  {\bibfnamefont {W.~D.}\ \bibnamefont {Oliver}}, \bibinfo {author}
  {\bibfnamefont {S.~O.}\ \bibnamefont {Valenzuela}}, \ and\ \bibinfo {author}
  {\bibfnamefont {T.~P.}\ \bibnamefont {Orlando}},\ }\href {\doibase
  10.1103/PhysRevLett.101.190502} {\bibfield  {journal} {\bibinfo  {journal}
  {Phys. Rev. Lett.}\ }\textbf {\bibinfo {volume} {101}},\ \bibinfo {pages}
  {190502} (\bibinfo {year} {2008})}\BibitemShut {NoStop}%
\bibitem [{\citenamefont {Izmalkov}\ \emph {et~al.}(2008)\citenamefont
  {Izmalkov}, \citenamefont {van~der Ploeg}, \citenamefont {Shevchenko},
  \citenamefont {Grajcar}, \citenamefont {Il'ichev}, \citenamefont {H\"ubner},
  \citenamefont {Omelyanchouk},\ and\ \citenamefont {Meyer}}]{izmalkov2}%
  \BibitemOpen
  \bibfield  {author} {\bibinfo {author} {\bibfnamefont {A.}~\bibnamefont
  {Izmalkov}}, \bibinfo {author} {\bibfnamefont {S.~H.~W.}\ \bibnamefont
  {van~der Ploeg}}, \bibinfo {author} {\bibfnamefont {S.~N.}\ \bibnamefont
  {Shevchenko}}, \bibinfo {author} {\bibfnamefont {M.}~\bibnamefont {Grajcar}},
  \bibinfo {author} {\bibfnamefont {E.}~\bibnamefont {Il'ichev}}, \bibinfo
  {author} {\bibfnamefont {U.}~\bibnamefont {H\"ubner}}, \bibinfo {author}
  {\bibfnamefont {A.~N.}\ \bibnamefont {Omelyanchouk}}, \ and\ \bibinfo
  {author} {\bibfnamefont {H.-G.}\ \bibnamefont {Meyer}},\ }\href {\doibase
  10.1103/PhysRevLett.101.017003} {\bibfield  {journal} {\bibinfo  {journal}
  {Phys. Rev. Lett.}\ }\textbf {\bibinfo {volume} {101}},\ \bibinfo {pages}
  {017003} (\bibinfo {year} {2008})}\BibitemShut {NoStop}%
\bibitem [{\citenamefont {Ferr\'on}\ \emph {et~al.}(2010)\citenamefont
  {Ferr\'on}, \citenamefont {Dom\'{\i}nguez},\ and\ \citenamefont
  {S\'anchez}}]{ferron2010}%
  \BibitemOpen
  \bibfield  {author} {\bibinfo {author} {\bibfnamefont {A.}~\bibnamefont
  {Ferr\'on}}, \bibinfo {author} {\bibfnamefont {D.}~\bibnamefont
  {Dom\'{\i}nguez}}, \ and\ \bibinfo {author} {\bibfnamefont {M.~J.}\
  \bibnamefont {S\'anchez}},\ }\href {\doibase 10.1103/PhysRevB.82.134522}
  {\bibfield  {journal} {\bibinfo  {journal} {Phys. Rev. B}\ }\textbf {\bibinfo
  {volume} {82}},\ \bibinfo {pages} {134522} (\bibinfo {year}
  {2010})}\BibitemShut {NoStop}%
\bibitem [{\citenamefont {Oliver}\ and\ \citenamefont
  {Valenzuela}(2009)}]{olivervalenzuela}%
  \BibitemOpen
  \bibfield  {author} {\bibinfo {author} {\bibfnamefont {W.~D.}\ \bibnamefont
  {Oliver}}\ and\ \bibinfo {author} {\bibfnamefont {S.~O.}\ \bibnamefont
  {Valenzuela}},\ }\href {\doibase 10.1007/s11128-009-0108-y} {\bibfield
  {journal} {\bibinfo  {journal} {Quantum Information Processing}\ }\textbf
  {\bibinfo {volume} {8}},\ \bibinfo {pages} {261} (\bibinfo {year}
  {2009})}\BibitemShut {NoStop}%
\bibitem [{\citenamefont {Shirley}(1965)}]{shirley}%
  \BibitemOpen
  \bibfield  {author} {\bibinfo {author} {\bibfnamefont {J.~H.}\ \bibnamefont
  {Shirley}},\ }\href {\doibase 10.1103/PhysRev.138.B979} {\bibfield  {journal}
  {\bibinfo  {journal} {Phys. Rev.}\ }\textbf {\bibinfo {volume} {138}},\
  \bibinfo {pages} {B979} (\bibinfo {year} {1965})}\BibitemShut {NoStop}%
\bibitem [{\citenamefont {Ferr\'on}\ \emph {et~al.}(2012)\citenamefont
  {Ferr\'on}, \citenamefont {Dom\'{\i}nguez},\ and\ \citenamefont
  {S\'anchez}}]{ferronprl}%
  \BibitemOpen
  \bibfield  {author} {\bibinfo {author} {\bibfnamefont {A.}~\bibnamefont
  {Ferr\'on}}, \bibinfo {author} {\bibfnamefont {D.}~\bibnamefont
  {Dom\'{\i}nguez}}, \ and\ \bibinfo {author} {\bibfnamefont {M.~J.}\
  \bibnamefont {S\'anchez}},\ }\href {\doibase 10.1103/PhysRevLett.109.237005}
  {\bibfield  {journal} {\bibinfo  {journal} {Phys. Rev. Lett.}\ }\textbf
  {\bibinfo {volume} {109}},\ \bibinfo {pages} {237005} (\bibinfo {year}
  {2012})}\BibitemShut {NoStop}%
\bibitem [{\citenamefont {Ferr\'on}\ \emph {et~al.}(2016)\citenamefont
  {Ferr\'on}, \citenamefont {Dom\'{\i}nguez},\ and\ \citenamefont
  {S\'anchez}}]{ferron2016}%
  \BibitemOpen
  \bibfield  {author} {\bibinfo {author} {\bibfnamefont {A.}~\bibnamefont
  {Ferr\'on}}, \bibinfo {author} {\bibfnamefont {D.}~\bibnamefont
  {Dom\'{\i}nguez}}, \ and\ \bibinfo {author} {\bibfnamefont {M.~J.}\
  \bibnamefont {S\'anchez}},\ }\href {\doibase 10.1103/PhysRevB.93.064521}
  {\bibfield  {journal} {\bibinfo  {journal} {Phys. Rev. B}\ }\textbf {\bibinfo
  {volume} {93}},\ \bibinfo {pages} {064521} (\bibinfo {year}
  {2016})}\BibitemShut {NoStop}%
\bibitem [{\citenamefont {Sauer}\ \emph {et~al.}(2012)\citenamefont {Sauer},
  \citenamefont {Mintert}, \citenamefont {Gneiting},\ and\ \citenamefont
  {Buchleitner}}]{sauer}%
  \BibitemOpen
  \bibfield  {author} {\bibinfo {author} {\bibfnamefont {S.}~\bibnamefont
  {Sauer}}, \bibinfo {author} {\bibfnamefont {F.}~\bibnamefont {Mintert}},
  \bibinfo {author} {\bibfnamefont {C.}~\bibnamefont {Gneiting}}, \ and\
  \bibinfo {author} {\bibfnamefont {A.}~\bibnamefont {Buchleitner}},\ }\href
  {http://stacks.iop.org/0953-4075/45/i=15/a=154011} {\bibfield  {journal}
  {\bibinfo  {journal} {Journal of Physics B: Atomic, Molecular and Optical
  Physics}\ }\textbf {\bibinfo {volume} {45}},\ \bibinfo {pages} {154011}
  (\bibinfo {year} {2012})}\BibitemShut {NoStop}%
\bibitem [{\citenamefont {Son}\ \emph {et~al.}(2009)\citenamefont {Son},
  \citenamefont {Han},\ and\ \citenamefont {Chu}}]{sang}%
  \BibitemOpen
  \bibfield  {author} {\bibinfo {author} {\bibfnamefont {S.-K.}\ \bibnamefont
  {Son}}, \bibinfo {author} {\bibfnamefont {S.}~\bibnamefont {Han}}, \ and\
  \bibinfo {author} {\bibfnamefont {S.-I.}\ \bibnamefont {Chu}},\ }\href
  {\doibase 10.1103/PhysRevA.79.032301} {\bibfield  {journal} {\bibinfo
  {journal} {Phys. Rev. A}\ }\textbf {\bibinfo {volume} {79}},\ \bibinfo
  {pages} {032301} (\bibinfo {year} {2009})}\BibitemShut {NoStop}%
\bibitem [{\citenamefont {Hausinger}\ and\ \citenamefont
  {Grifoni}(2010)}]{griffoni}%
  \BibitemOpen
  \bibfield  {author} {\bibinfo {author} {\bibfnamefont {J.}~\bibnamefont
  {Hausinger}}\ and\ \bibinfo {author} {\bibfnamefont {M.}~\bibnamefont
  {Grifoni}},\ }\href {\doibase 10.1103/PhysRevA.81.022117} {\bibfield
  {journal} {\bibinfo  {journal} {Phys. Rev. A}\ }\textbf {\bibinfo {volume}
  {81}},\ \bibinfo {pages} {022117} (\bibinfo {year} {2010})}\BibitemShut
  {NoStop}%
\bibitem [{\citenamefont {Ashhab}\ \emph {et~al.}(2007)\citenamefont {Ashhab},
  \citenamefont {Johansson}, \citenamefont {Zagoskin},\ and\ \citenamefont
  {Nori}}]{nori2}%
  \BibitemOpen
  \bibfield  {author} {\bibinfo {author} {\bibfnamefont {S.}~\bibnamefont
  {Ashhab}}, \bibinfo {author} {\bibfnamefont {J.~R.}\ \bibnamefont
  {Johansson}}, \bibinfo {author} {\bibfnamefont {A.~M.}\ \bibnamefont
  {Zagoskin}}, \ and\ \bibinfo {author} {\bibfnamefont {F.}~\bibnamefont
  {Nori}},\ }\href {\doibase 10.1103/PhysRevA.75.063414} {\bibfield  {journal}
  {\bibinfo  {journal} {Phys. Rev. A}\ }\textbf {\bibinfo {volume} {75}},\
  \bibinfo {pages} {063414} (\bibinfo {year} {2007})}\BibitemShut {NoStop}%
\bibitem [{\citenamefont {Wootters}(1998)}]{wootters}%
  \BibitemOpen
  \bibfield  {author} {\bibinfo {author} {\bibfnamefont {W.~K.}\ \bibnamefont
  {Wootters}},\ }\href {\doibase 10.1103/PhysRevLett.80.2245} {\bibfield
  {journal} {\bibinfo  {journal} {Phys. Rev. Lett.}\ }\textbf {\bibinfo
  {volume} {80}},\ \bibinfo {pages} {2245} (\bibinfo {year}
  {1998})}\BibitemShut {NoStop}%
\bibitem [{\citenamefont {Chow}\ \emph {et~al.}(2010)\citenamefont {Chow},
  \citenamefont {DiCarlo}, \citenamefont {Gambetta}, \citenamefont
  {Nunnenkamp}, \citenamefont {Bishop}, \citenamefont {Frunzio}, \citenamefont
  {Devoret}, \citenamefont {Girvin},\ and\ \citenamefont
  {Schoelkopf}}]{dicarlo}%
  \BibitemOpen
  \bibfield  {author} {\bibinfo {author} {\bibfnamefont {J.~M.}\ \bibnamefont
  {Chow}}, \bibinfo {author} {\bibfnamefont {L.}~\bibnamefont {DiCarlo}},
  \bibinfo {author} {\bibfnamefont {J.~M.}\ \bibnamefont {Gambetta}}, \bibinfo
  {author} {\bibfnamefont {A.}~\bibnamefont {Nunnenkamp}}, \bibinfo {author}
  {\bibfnamefont {L.~S.}\ \bibnamefont {Bishop}}, \bibinfo {author}
  {\bibfnamefont {L.}~\bibnamefont {Frunzio}}, \bibinfo {author} {\bibfnamefont
  {M.~H.}\ \bibnamefont {Devoret}}, \bibinfo {author} {\bibfnamefont {S.~M.}\
  \bibnamefont {Girvin}}, \ and\ \bibinfo {author} {\bibfnamefont {R.~J.}\
  \bibnamefont {Schoelkopf}},\ }\href {\doibase 10.1103/PhysRevA.81.062325}
  {\bibfield  {journal} {\bibinfo  {journal} {Phys. Rev. A}\ }\textbf {\bibinfo
  {volume} {81}},\ \bibinfo {pages} {062325} (\bibinfo {year}
  {2010})}\BibitemShut {NoStop}%
\bibitem [{\citenamefont {Shulman}\ \emph {et~al.}(2012)\citenamefont
  {Shulman}, \citenamefont {Dial}, \citenamefont {Harvey}, \citenamefont
  {Bluhm}, \citenamefont {Umansky},\ and\ \citenamefont {Yacoby}}]{shulman202}%
  \BibitemOpen
  \bibfield  {author} {\bibinfo {author} {\bibfnamefont {M.~D.}\ \bibnamefont
  {Shulman}}, \bibinfo {author} {\bibfnamefont {O.~E.}\ \bibnamefont {Dial}},
  \bibinfo {author} {\bibfnamefont {S.~P.}\ \bibnamefont {Harvey}}, \bibinfo
  {author} {\bibfnamefont {H.}~\bibnamefont {Bluhm}}, \bibinfo {author}
  {\bibfnamefont {V.}~\bibnamefont {Umansky}}, \ and\ \bibinfo {author}
  {\bibfnamefont {A.}~\bibnamefont {Yacoby}},\ }\href {\doibase
  10.1126/science.1217692} {\bibfield  {journal} {\bibinfo  {journal}
  {Science}\ }\textbf {\bibinfo {volume} {336}},\ \bibinfo {pages} {202}
  (\bibinfo {year} {2012})},\ \Eprint
  {http://arxiv.org/abs/http://science.sciencemag.org/content/336/6078/202.full.pdf}
  {http://science.sciencemag.org/content/336/6078/202.full.pdf} \BibitemShut
  {NoStop}%
\bibitem [{\citenamefont {Romero}\ \emph {et~al.}(2007)\citenamefont {Romero},
  \citenamefont {L\'opez}, \citenamefont {Lastra}, \citenamefont {Solano},\
  and\ \citenamefont {Retamal}}]{romero}%
  \BibitemOpen
  \bibfield  {author} {\bibinfo {author} {\bibfnamefont {G.}~\bibnamefont
  {Romero}}, \bibinfo {author} {\bibfnamefont {C.~E.}\ \bibnamefont {L\'opez}},
  \bibinfo {author} {\bibfnamefont {F.}~\bibnamefont {Lastra}}, \bibinfo
  {author} {\bibfnamefont {E.}~\bibnamefont {Solano}}, \ and\ \bibinfo {author}
  {\bibfnamefont {J.~C.}\ \bibnamefont {Retamal}},\ }\href {\doibase
  10.1103/PhysRevA.75.032303} {\bibfield  {journal} {\bibinfo  {journal} {Phys.
  Rev. A}\ }\textbf {\bibinfo {volume} {75}},\ \bibinfo {pages} {032303}
  (\bibinfo {year} {2007})}\BibitemShut {NoStop}%
\bibitem [{\citenamefont {Akisato}(2008)}]{akisato}%
  \BibitemOpen
  \bibfield  {author} {\bibinfo {author} {\bibfnamefont {H.}~\bibnamefont
  {Akisato}},\ }\href {arXiv:quant-ph/0303128} {\  (\bibinfo {year}
  {2008})}\BibitemShut {NoStop}%
\bibitem [{\citenamefont {Satanin}\ \emph {et~al.}(2012)\citenamefont
  {Satanin}, \citenamefont {Denisenko}, \citenamefont {Ashhab},\ and\
  \citenamefont {Nori}}]{satanin}%
  \BibitemOpen
  \bibfield  {author} {\bibinfo {author} {\bibfnamefont {A.~M.}\ \bibnamefont
  {Satanin}}, \bibinfo {author} {\bibfnamefont {M.~V.}\ \bibnamefont
  {Denisenko}}, \bibinfo {author} {\bibfnamefont {S.}~\bibnamefont {Ashhab}}, \
  and\ \bibinfo {author} {\bibfnamefont {F.}~\bibnamefont {Nori}},\ }\href
  {\doibase 10.1103/PhysRevB.85.184524} {\bibfield  {journal} {\bibinfo
  {journal} {Phys. Rev. B}\ }\textbf {\bibinfo {volume} {85}},\ \bibinfo
  {pages} {184524} (\bibinfo {year} {2012})}\BibitemShut {NoStop}%
\end{thebibliography}%
\end{document}